\documentclass{aa}  

\usepackage{graphicx}
\usepackage{txfonts}
\usepackage{xcolor}

\newcommand\kms{{\rm\,km\,s^{-1}}}
\newcommand\msun{\rm\,M_\odot}

 \def\simle{\mathrel{\hbox{\rlap{\hbox{\lower4pt\hbox{$\sim$}}}\hbox{$<$}}}}
 \def\simgr{\mathrel{\hbox{\rlap{\hbox{\lower4pt\hbox{$\sim$}}}\hbox{$>$}}}}

\begin{document}

   \title{Unequal-mass, highly-spinning binary black hole mergers in the stable mass transfer formation channel}

   \author{Aleksandra Olejak
         \inst{1}
         \and
          Jakub Klencki
          \inst{2}
          \and
          Xiao-Tian Xu
          \inst{3}
          \and
          Chen Wang,
          \inst{1}
          {\textdagger Krzysztof Belczynski}
          \inst{4}
          \and
          Jean-Pierre Lasota
          \inst{4,}
          \inst{5}
          }

   \institute{Max Planck Institut für Astrophysik, Karl-Schwarzschild-Stra{\ss}e 1,  85748 Garching bei München, Germany\\
              \email{aolejak@mpa-garching.mpg.de}
        \and
             European Southern Observatory, Karl-Schwarzschild-Stra{\ss}e 2, 85748, Garching bei München, Germany
        \and
             Argelander-Institut f\"ur Astronomie, Universit\"at Bonn, Auf dem H\"ugel 71, 53121 Bonn, Germany
        \and
             Nicolaus Copernicus Astronomical Center of Polish Academy of Sciences, Bartycka 18, 00-716 Warszawa, Poland
        \and
             Institut d’Astrophysique de Paris, CNRS et Sorbonne Université, UMR 7095, 98bis Bd Arago, 75014 Paris, France
             }

   \date{Received April, 2024 ; accepted ...}

 
  \abstract
   {The growing database of gravitational-wave (GW) detections with the binary black holes (BHs) merging in the distant Universe contains subtle insights into their formation scenarios.}
   {We investigate one of the puzzling properties of detected GW sources which is the possible (anti)correlation between mass ratio $q$ of BH-BH binaries and their effective spin $\chi_{\rm eff}$. In particular, unequal-mass systems tend to exhibit higher spins than the ones with nearly equal-mass BH components.}
   {We use rapid binary evolution models to demonstrate that the isolated binary evolution followed by efficient tidal spin-up of stripped helium core produces a similar pattern in $\chi_{\rm eff}$ vs $q$ distributions of BH-BH mergers.}
   {In our models, the progenitors of unequal BH-BH systems in the stable mass transfer formation scenario are more likely to efficiently shrink their orbits during the second Roche-lobe overflow than the binaries that evolve into nearly equal-mass component systems. This makes it easier for unequal-mass progenitors to enter the tidal spin-up regime and later merge due to GW emission. Our results are, however, sensitive to some input assumptions, especially, the stability of mass transfer and the angular momentum loss during non-conservative mass transfer. We note that mass transfer prescriptions widely adopted in rapid codes favor the formation of BH-BH merger progenitors with unequal masses and moderate separations. We compare our results with detailed stellar model grids and find reasonable agreement after appropriate calibration of the physics models. }
   {We anticipate that future detections of unequal-mass BH-BH mergers could provide valuable constraints on the role of the stable mass transfer formation channel. A significant fraction of BH-BH detections with mass ratio $q \in (0.4-0.7)$ would be consistent with the mass ratio reversal scenario during the first, relatively conservative mass transfer, and a non-enhanced angular momentum loss during the second, highly non-conservative mass transfer phase. }


   \maketitle
%

\section{Introduction}

Recent analysis of detected gravitational wave (GW) sources reports on the possible negative correlation between mass ratios $q$ (defined as the mass of the less massive over the more massive object) and effective spin parameter $\chi_{\rm eff}$\footnote{$\chi_{\rm eff}= \frac{m_1 {\chi}_{\rm 1} \cos \theta_1+m_2 {\chi}_{\rm 2} \cos \theta_2}{m_1 + m_2}$
where $m_{i}$ are BH masses, ${\chi}_{i}=cJ_{i}/Gm_{i}^2$ are dimensionless spin magnitudes of BHs, 
$\theta_{i}$ are angles between the individual BH spins and the system's orbital angular momentum, $c$ is the speed of the light in the vacuum, and $G$ is the gravitational constant. } among binary black hole mergers (BH-BH) announced by LIGO-Virgo-Kagra (LVK) collaboration \citep{LIGOfullO3population2021, Callister2021, Adamcewicz2023}. In particular, unequal-mass BH-BH systems tend to have higher inferred effective spin values than the mergers characterized by equal-mass BH components. If this correlation is real, (i.e., not being the result of possible degeneracies between individual parameter measurements, see e.g., \citealt{Hannam2013} and \citealt{Mandel2021a})
it would indicate that a specific mechanism is at work during the formation of at least some current GW sources. Exploring the possibility of such correlations between $\chi_{\rm eff}$ and $q$ by modeling properties of BH-BH mergers populations from different formation channels may help in constraining their relative contributions and put constraints on the uncertain astrophysical processes. \cite{McKernan2022}, \cite{Vaccaro2023}  and \cite{Xeff_agndisk2023} obtained a similar trend in $q - \chi_{\rm eff}$ distribution in their models once they combined the synthetic populations of isolated field binaries with a contribution of hierarchical mergers produced in active galactic nuclei environment.

In this paper, we present an alternative possibility of reproducing the inferred $q - \chi_{\rm eff}$ trend by modeling only isolated binary progenitors of GW sources \citep{LIGOfullO3population2021}. In particular, we demonstrate that the contribution of BH-BH mergers formed via stable mass transfer and common envelope (CE) subchannels may also result in a characteristic pattern in $q - \chi_{\rm eff}$ distribution, similar to the one reported for LVK sources. 

The classical isolated binary formation scenario for BH-BH mergers, which has been popular in literature for several years, includes a CE phase \citep{Belczynski2016b,Eldridge2016,Stevenson2017,Kruckow2018,Hainich2018,Marchant2018,Spera2019,Mapelli2019,Bavera2020}. This scenario consists of a stable mass transfer phase during the first Roche lobe overflow (RLOF) and a CE (i.e., dynamically unstable mass transfer phase) during the second RLOF. The CE is considered a promising mechanism in binary evolution for bringing an initially wide system close enough to lead to a double compact object merger \citep{Paczynski1976}.
However, recently, the significant contribution of the CE scenario has been challenged for a few independent reasons. First, several studies indicate that mass transfer in massive binary systems, such as BH-BH progenitors, is more stable compared to what has been previously found \citep{Ge2010, Ge2015, Pavlovskii2017, Ge2020b, Ge2020a, Marchant2021,Shao2021}. Second, even if unstable mass transfer develops in a system, successful envelope ejection can only happen under very restrictive conditions, making the merger of a donor star and a BH a likely outcome \citep{Kruckow2016, Klencki2020, Marchant2021}. As a result, the merger rates predicted for BH-BH systems formed via CE scenario could be significantly overestimated \citep{Gallegos-Garcia2021}. 

During the last few years, the alternative isolated binary BH-BH formation scenario consisting of a stable mass transfer phase during the second RLOF (instead of CE) has been gaining popularity in the GW community \citep[see e.g.,][]{vandelHeuvel2017,Neijssel2019,Marchant2021,Bavera2021,Olejak2021a,vanSon2021,Olejak2021b,Shao2021,Broekgaarden2022,vanSon2022,Briel2023,Dorozsmai2024,Picco2024}. 
Stable and unstable mass transfer are expected to proceed on specific time scales and differ in the way that masses and orbit evolve. Therefore, the mass transfer stability determines the final fate of massive binary systems and impacts the distribution of final masses, orbital parameters, spins, and merger rates of BH-BH merger population \citep{Olejak2021a,vanSon2021,Bavera2022,Olejak2021b,vanSon2022,Broekgaarden2022,Dorozsmai2024}. Once allowing for relatively conservative mass transfer during the first RLOF, the mass ratio of the binary system may get reversed, so the secondary (initially less massive) star would become a few times more massive than the primary (initially more massive) star \citep{Olejak2021b,Broekgaarden2022}. If such a significant mass ratio reversal during the first RLOF (to $M_{\rm comp}/M_{\rm don} \gtrsim 3$ at its end) is common, it would favor the formation of unequal-mass BH-BH binaries.
Note that some recent analyses of LVK detections find support for the contribution of such a mass-ratio reversal formation scenario \citep{Broekgaarden2022,Adamcewicz2023,Adamcewicz2024} as well as unequal-mass BH-BH mergers \citep{LIGOfullO3population2021,Rinaldi2023,Sadiq2023}.

So far, LVK analysis of the detected BH-BH population is consistent with distribution dominated by nearly equal-mass component mergers \citep{LIGO2019a}. However, some recent works that adopt alternative approaches to infer GW source parameters, find evidence or even a preference for the contribution of unequal-mass ratio BH-BH mergers \citep{Rinaldi2023,Sadiq2023}. In particular, they found GW data with BH-BH consistent with the peak in the mass ratio between 0.4-0.6, aligning with the distribution derived from our stable mass transfer scenario followed by mass ratio reversal. The mass ratio of BH-BH mergers inferred from GW data is, however, uncertain. The general properties extracted for the BH-BH mergers population such as the distribution of mass ratio (or other parameters) are sensitive to the adopted prior assumptions \citep{Farah2024}. 

A stable mass transfer channel in our models may result in a significant fraction of highly spinning, unequal-mass BH-BH mergers \citep{Olejak2021b}. This is in contrast to some other recent studies, which found that stable mass scenarios are rather unlikely to reproduce detected BH-BH systems with non-negligible $\chi_{\rm eff}>0$ \citep{Zevin2022} unless adopting highly super-Eddington accretion on a BH. In this paper, we focus on the properties of BH-BH progenitors in stable mass transfer and CE subchannels. We explain why our models, which differ from most of the similar rapid population synthesis codes by criteria for mass transfer stability, favor the formation of unequal-mass BH-BH mergers with $q \in(0.4-0.7)$ and allow for a significant fraction of highly spinning systems. We also test how the properties of produced BH-BH mergers are sensitive to uncertain assumptions on angular momentum loss, BH natal kicks, and mass transfer efficiency.

The paper is organized as follows: in Section \ref{sec:method} we describe the method and the relevant physical assumptions adopted in our models. Section \ref{sec: results} is devoted to progenitors of BH-BH mergers and understanding the role of the angular momentum loss (Sec. \ref{Sec: ang_mom_loss}). Section \ref{sec: time_delays} presents the time delay and mass ratio distribution of BH-BH mergers in different tested physical models. Section \ref{sec: evolutionary_scenario} includes evolutionary scenario schema and Section \ref{sec:natal_kicks} addresses the role of natal kicks. In Section \ref{sec:discussion} we discuss other uncertainties and the weak points of modeling binary evolution using rapid population synthesis codes that could affect our results. The discussion includes a brief comparison with BH-star binaries grids generated with detailed stellar evolution codes. In Section \ref{sec:conclusions} we provide a summary and conclusions. In the appendix, we address the role of mass transfer instability in tight mass transferring binaries (Appendix \ref{Sec: exp_instab})

\section{Method} \label{sec:method}

We use the {\tt StarTrack} population synthesis code \citep{Belczynski2008,Belczynski2020}  with a few recent updates \citep{Belczynski2020PSN,Olejak2021a,Olejak2022} to generate a population of merging BH-BH systems. To evaluate merger rate density as a function of redshift $z$, synthetic BH-BH systems are post-processed using models of star formation rates and metallicity distribution evolution in the Universe by \cite{MadauFragos2017} as described in \cite{Dominik2015} and \cite{Belczynski2020}. Below, we address a few subjectively selected physical assumptions most relevant to this study.

Our default model adopts revised mass transfer stability criteria based on results of \cite{Pavlovskii2017}. They revisited mass transfer stability for a grid of massive BH-star systems (possible BH-BH progenitors) allowing for high degrees of the RLOF (to the outer Lagrangian point of the donor) and high mass transfer rates (of a few percent of the dynamical timescale). They found that mass transfer in such types of binaries is significantly more stable than was previously expected. The revised criteria, as implemented in \cite{Olejak2021a}, significantly limit parameter space for CE (i.e., unstable mass transfer) development in comparison to the formerly used one \citep{Belczynski2008}. They are characterized by more strictly limited conditions for the mass ratio of the donor to the accertor, which before typically was $\sim 2-3$ \citep{Belczynski2008} and now increased to $\sim 3-5$ \citep{Olejak2021a}. The new criteria also include extra conditions for donor evolutionary type, radii, and mass to develop unstable mass transfer. The revised criteria have different variants for high (Z$>0.01$) and low (Z$\leq0.01$) metallicities. As a result, even binaries with highly unequal masses, such that the donor is $\sim 6-7$ times more massive than the accretor, may remain stable during the mass transfer phase depending on the individual properties of the system. The default version of our criteria, however, takes into account extra parameter space for unstable mass transfer that could develop in tight binaries found, e.g., by \cite{Pavlovskii2017}, described in Appendix \ref{Sec: exp_instab}). The full and detailed description of our mass transfer stability criteria can be found in Section 3.1 of \cite{Olejak2021a}

For mass transfer onto a non-degenerate accretor, we assume a fixed accretion efficiency $\beta = 50\%$, where $\beta$ is a fraction of transferred mass accreted by the companion star, motivated by e.g. \cite{Vinciguerra2020}. The non-accreted matter is lost from the system with the specific angular momentum of the binary \citep{Podsiadlowski1992}. The possible accretion rate is, however, assumed to be limited by the Eddington rate corresponding to the radius of the accretor.\footnote{The calculated Eddington rate does not take into account the rapid increase in the total accretor radius expected to happen in case of accretion during the mass transfer on short (thermal and below) timescales, see \cite{Lau2024,Schurmann2024}.}$^{,}$\footnote{That limit affects mainly progenitors of massive BH-BH mergers in CE subchannel with wide initial orbital separations of $a \gtrsim 800 R_{\odot}$.} In the case of accretion onto a BH component, we adopt the analytic approximations \citep{King2001} implemented by \cite{Mondal2020}. This approach results in a highly non-conservative mass transfer from the companion to the BH, with the accretion rate onto the BH equal to the  Eddington rate \citep[see e.g.][]{King2023}. The non-accreted mass in our default model is lost with the specific angular momentum of the accretor. This assumption has been, however, questioned by e.g., \cite{Gallego0823} who pointed out that the real angular momentum loss might be much higher once taking into account e.g., winds from an accretion disk around the compact objected accretor.
Moreover, the presence of a circumbinary disk which could possibly form after a rapid, non-conservative mass transfer phase, could affect the orbital evolution due to interaction between disk and binary, and a possible angular momentum exchange \citep{Pejcha2016,Dorazio2021,Zrake2021,Siwek2023,Valli2024}. 
Therefore, in our study, we also test a model with a significantly higher angular momentum loss for which the material is lost with the specific angular momentum loss of the outer Lagrange point. In our models, we assume efficient circularization and synchronization due to tides, so during the RLOF systems are always on circular orbits. Even if at the onset of the RLOF phase, the orbit is still eccentric, once the mass transfer begins the system separation is circularized to the periastron \citep{Belczynski2008}. 

We adopt the so-called {\textit{rapid}} type supernovae (SN) engine, with the mixing parameter values $f_{\rm mix}=2.5$ of convection-enhanced supernova engines by \cite{Fryer2022}. We assume that BH formation is accompanied by a natal kick. In our default model, kicks are derived from a Maxwellian velocity distribution with $\sigma = 265 \kms$ \citep{Hobbs2005}, however, their magnitude is decreased by the amount of fallback according to prescriptions by \cite{Fryer2012} and \cite{Belczynski2012}. Such a prescription makes massive BHs unlikely to get significant natal kicks. BH natal kicks are usually expected to be much smaller than the natal kicks of neutron stars \citep{Janka2024}. However, so far the constraints based on the fits to observed systems  \citep{Jonker2004, Repetto2012, Cesares2014,Repetto2017, VignaGomez2023} are rather poor. Some theoretical studies indicate that compact object formation might be followed by two independent types of kicks that originate either from asymmetric mass ejection or emission of neutrinos, e.g., \cite{Fryer2006} and \cite{Janka2024}. Due to the unconstrained nature of BH natal kicks in a few cases in addition to our default model, we test an alternative natal kick model with high BH kicks. In that model, the kick velocity is not decreased by fallback. Instead, we reduce the $\sigma$ of Maxwellian velocity distribution by a factor of two to $\sigma= 133 \kms$. Note that this model results in high BH natal kicks, much larger than suggested by some recent theoretical predictions \citep[e.g.,][]{Janka2024}.

Motivated by several massive BH-BH mergers detections (with $M_{\rm BH}$$\geq 50 M_{\odot}$), we adopt a high limit for pair-instability supernovae (PSN), assuming that stars with their final helium core masses above $M_{\rm He}>90 M_{\odot}$ get disrupted \citep{Belczynski2020PSN}. The high PSN limit is also justified by significant uncertainties in $^{12}$C({\ensuremath{\alpha}}, {\ensuremath{\gamma}})$^{16}$O reaction rate \citep{Farmer2020,2021Costa, Woosley2021,Farag2022,Hendriks2023}. We note, however, that the adopted PSN limit does not noticeably affect any of the presented results, such as BH-BH merger rates, mass ratio, or formation scenarios, see e.g. \cite{Olejak2022}.
For natal BH spins we adopt low but non-zero, positive values $\chi \approx 0.05-0.15$ \citep{Belczynski2020} derived under the assumption of efficient angular momentum transport in massive stars \citep{Spruit_2002}. We adopt efficient tidal spin-up of stripped helium cores in close BH and stripped helium core systems with orbital periods below 1.1 days \citep{Izzard2004,Detmers2008, Kushnir2016, Hotokezaka2017, Zaldarriaga2018,Belczynski2020, Bavera2020}.

\section{BH-BH progenitors in the stable mass transfer channel}

\subsection{Stable mass transfer vs common envelope} \label{sec: results}

Evolution through stable mass transfer affects the binary orbital separation as well as its components in a different way than the CE phase, which later affects the properties of formed BH-BH mergers \citep[for some recent studies see e.g.][]{Olejak2022,Zevin2022,vanSon2022,Willcox2023}. Stable mass transfer is expected to produce in general wider BH-BH systems than CE, as the widely used by rapid population synthesis $\alpha$ prescription for the CE outcome tends to shrink the orbital separation more efficiently than the loss of angular momentum during non-conservative stable mass transfer. However, the actual outcome for an individual system depends on its properties at the onset of the stable mass transfer or CE phase, in particular the mass ratio between the donor and the accretor, and the orbital period \citep[see e.g.][]{deMarco2011,Ge2022,Ge2024}. Such a dependency produces characteristic fingerprints in the parameter distribution of the BH-BH systems subpopulation which merge within the age of Universe time.

\begin{figure*}[!htbp] 
\includegraphics[width=0.99\textwidth]{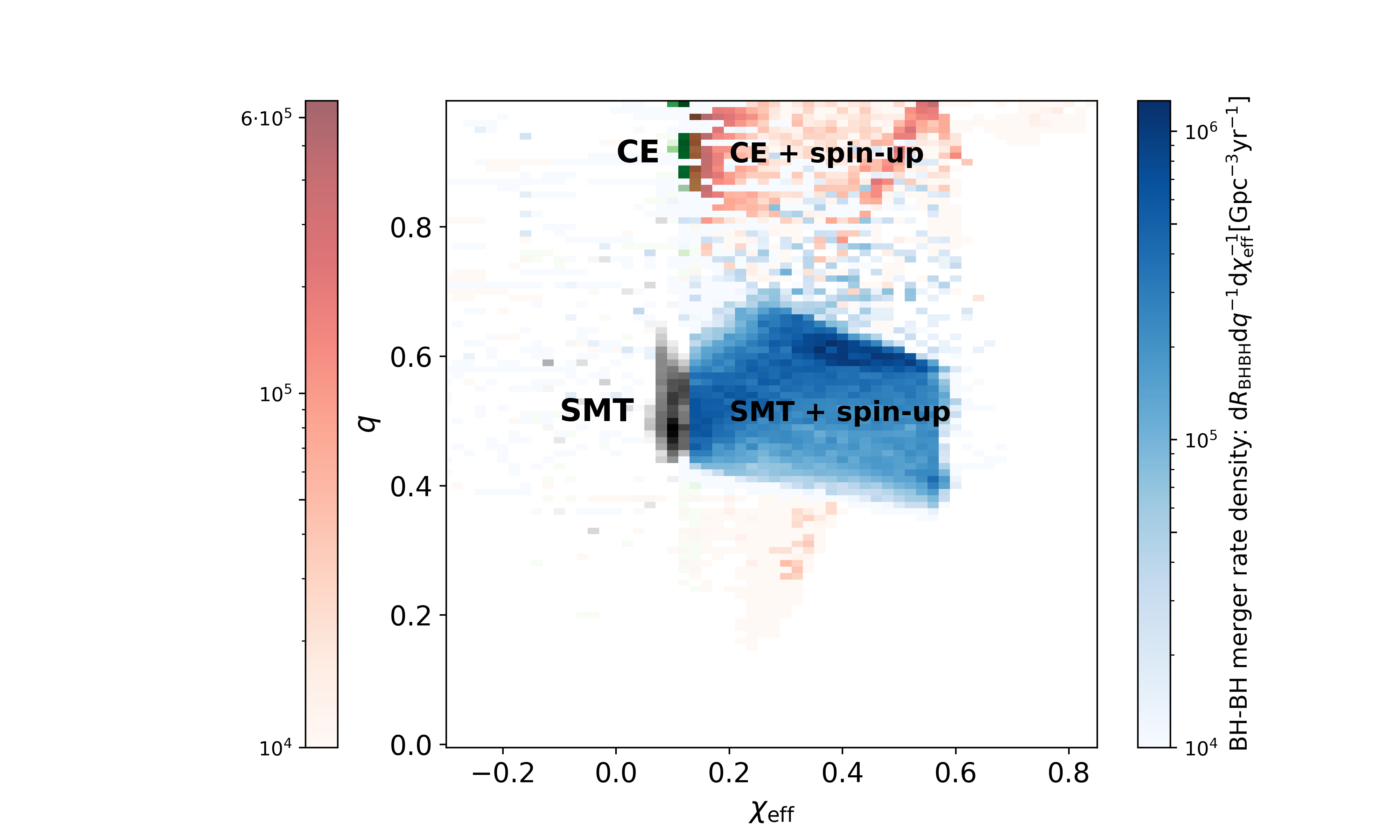}
\caption{Two-dimensional $\chi_{\rm eff}$ - $q$ histograms of BH-BH merger rate density $\frac{{\rm d}R_{\rm BHBH}}{{\rm d}q{\rm d} \chi_{\rm eff}}$[Gpc$^{-3}$yr$^{-1}$] up to redshifts $z_{\rm mer}<1.0$ in our default model with the revised mass transfer stability criteria \citep{Pavlovskii2017,Olejak2021a}. BH-BH mergers are produced via stable mass transfer (SMT, black, and blue colors) and common envelope (CE, green and red colors) subchannels. The formation of highly spinning mergers with $\chi_{\rm eff} \gtrapprox 0.15$  was followed by a tidal spin-up phase in close BH-helium core binaries (blue and red colors).}
\label{fig: Xeff_q_SMT_CE}
\end{figure*}

We find that the revised mass transfer stability criteria limiting CE development combined with low (non-enhanced) angular momentum loss during the second, stable mass transfer phase favor the formation of unequal-mass ratio BH-BH mergers with the characteristic, broad peak in the distribution between $q \in 0.4-0.7$ (see Sec. \ref{Sec: ang_mom_loss} and Figure \ref{fig: mass_ratio_three_models}). \footnote{Note that emergence of this feature was already reported by a few other recent studies using {\tt StarTrack} as well as other rapid population synthesis codes, once applying similar input assumptions \citep{Olejak2021a,vanSon2021,vanSon2022,Olejak2022,Broekgaarden2022,Dorozsmai2024}.} Moreover, the progenitors of those unequal BH-BH mergers may experience an efficient tidal spin-up phase at the later evolutionary phase, after which the stripped helium core can produce rapidly rotating second-born BH \citep{Detmers2008, Kushnir2016, Hotokezaka2017, Zaldarriaga2018,Belczynski2020, Bavera2020}. In our default physical model, within the same set of input assumptions, the CE formation channel is subdominant and tends to produce nearly equal-mass BH-BH merger with low, positive effective spin $\chi_{\rm eff}$ values. Combination of BH-BH mergers formed via the subchannels, with and without CE, results in characteristic $\chi_{\rm eff}$ and $q$ patterns in their distribution in our simulations, shown in Figure \ref{fig: Xeff_q_SMT_CE}. 

Our criteria for mass transfer stability limit the development of CE much more than typically assumed in widely used rapid population synthesis code. That explains why other studies did not find a similar formation scenario for highly spinning BH-BH mergers via a stable mass transfer channel. In particular, in other codes, highly unequal BH-star systems (the main BH-BH progenitors in our stable mass transfer channel) initiate an unstable mass transfer phase during the second RLOF. Therefore, in contrast to our results, \cite{Zevin2022} and \cite{Bavera2022} found a stable mass transfer channel rather unlikely to reproduce the unequal-mass, highly spinning BH-BH mergers. However, with a similar set of physical assumptions, the population of BH-BH mergers in our simulations resembles in terms of the mass and spin distribution prediction using other rapid population synthesis codes, see e.g., \cite{Broekgaarden2022,vanSon2022}.

\begin{figure*}[!htbp]
\includegraphics[width=0.99\textwidth]{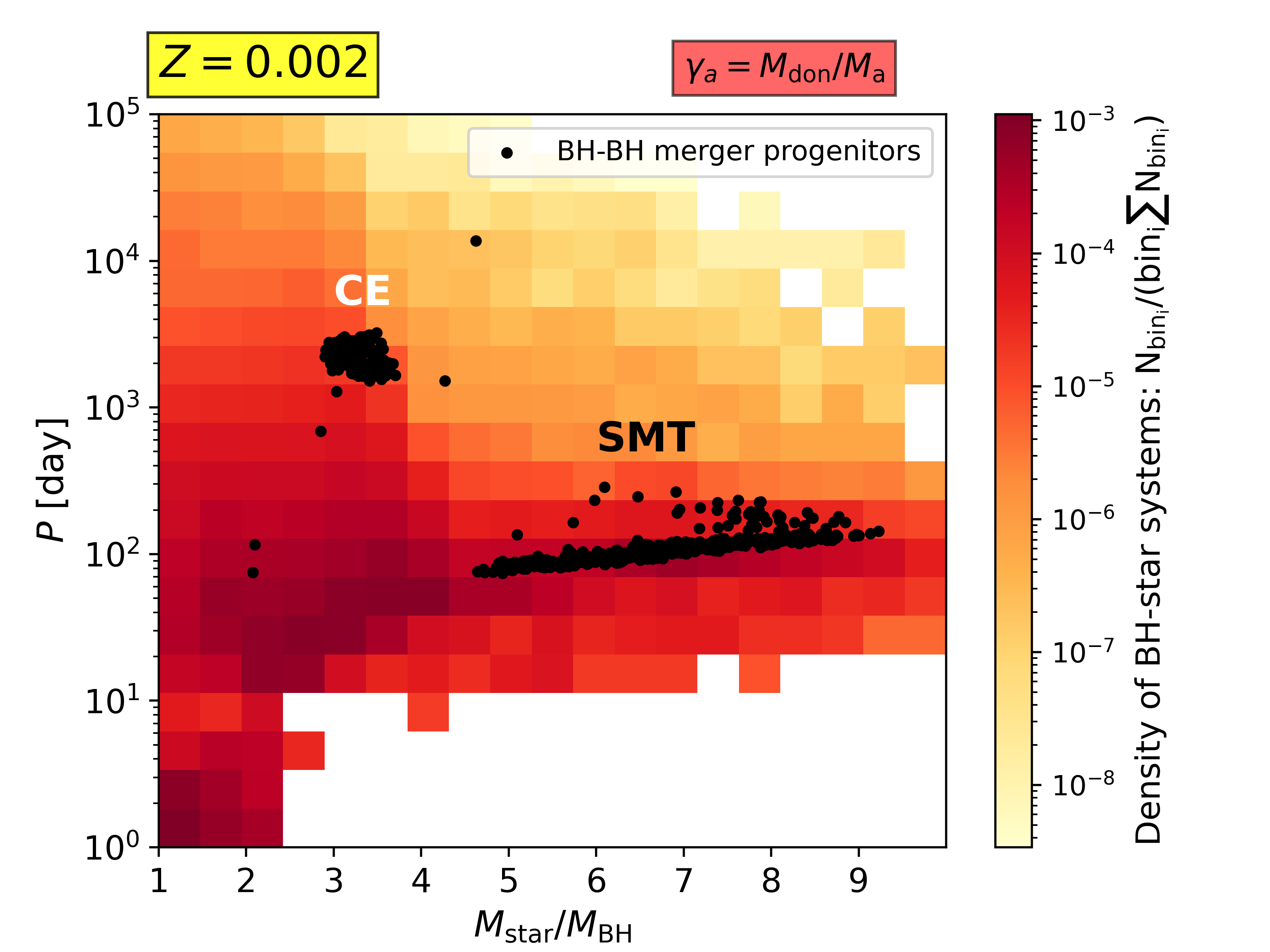} 
\caption{Orbital period vs mass ratio distribution of BH-star binaries that evolved from initially massive star systems, at the moment of the first BH formation (includes also the systems that evolved through CE). Results for our default physical model. The progenitors of BH-BH mergers via two formation scenarios: common envelope (CE) and stable mass transfer channel (SMT), are marked by the black dots.}
\label{fig: Period_q_SMT_and_CE}
\end{figure*}

In Figure \ref{fig: Period_q_SMT_and_CE} we present the density grid of BH-star binaries in terms of their orbital periods and mass ratios, at the moment of the first BH formation. This grid (in contrast to the similar figures presented in the next section \ref{Sec: ang_mom_loss}) includes also the BH-star binaries that evolved through the CE phase. The results are systems that evolved from 100,000 massive stellar binaries with their initial masses in the range $M_{1}>15 M_{\odot}$ and $M_{2}>$5 $M_{\odot}$ and with metallicity of 10\% solar ($Z=0.002$).
The progenitors of BH-BH mergers in the two subchannels: CE and SMT, are marked by the black dots.
The figure demonstrates that the progenitors of BH-BH mergers in the two subchannels favor specific separations and mass ratios at the moment of the first BH formation. Progenitors of BH-BH mergers in the CE channel have typically wide separations at the onset of the second RLOF and mass ratios $M_{\rm star}/M_{M_{\rm BH}} \approx 3-4$ - those systems can avoid a merger during the CE phase. The BH-BH mergers progenitors in SMT are characterized by more extreme mass ratios $M_{\rm star}/M_{M_{\rm BH}} > 4$ and moderate orbital periods of $P \approx 100$ days. The reason for such specific properties of BH-BH merger progenitors in SMT has been addressed in the next Section \ref{Sec: ang_mom_loss}.
Note, that this is a limited grid of 100,000 massive binaries we used to demonstrate the preferable mass ratios and separations of BH-BH progenitors for the two subchannels: stable mass transfer and CE in our default model. This model, due to e.g. relatively conservative mass transfer produces a significant fraction of unequal-mass BH-star binaries on moderate separations (read more in Sec. \ref{Sec: ang_mom_loss} and discussion in Sec. \ref{sec:discussion}). BH-BH mergers can, however, also evolve from different mass ratios and separations. For example, two black dots on the left side corresponding to $M_{\rm star}/M_{\rm BH} \approx 2$ will also later evolve in BH-BH mergers via stable mass transfer subchannel. The stars will initiate the second mass transfer on their core helium-burning phase, losing their envelope in mass transfer and finally evolving into BH-BH binaries with mass ratio $q \approx 0.5$ and a long time delay to the merger of several Gyr.

\subsection{The role of the angular momentum loss} \label{Sec: ang_mom_loss}

The first RLOF in our stable mass transfer scenario is relatively conservative. BH-BH progenitors in this subchannel with initial separation typically of the order of $\sim 100$ $R_{\odot}$, begin mass transfer early after the primary leaves its main sequence. The accretor (main sequence star) usually gains 50\% of the transferred companion mass (see Sec. \ref{sec:method}). In contrast, the second RLOF in our models with the accretion rate onto the BH equal to the  Eddington rate \citep{King2023} results in a highly non-conservative mass transfer with a vast majority of transferred mass lost from the system. 

The structure of the accretion flow to the BH accretor during the second mass transfer phase is uncertain. First, the formation of an accretion disk might depend on the BH spin \citep{Sen0821} which impacts the limit on the accretion rate \citep{Begelman0478}. Even if the disk forms, its geometry and mass loss from the disk are uncertain, with several different approaches suggested by the literature \citep{SS73,Abramowicz0988,Tetarenko0218,Yoshioka1222, Hu0822,King2023,Gallego0823}. 
The geometry of the accretion disk not only determines the amount of mass loss but also how the material is lost and therefore what is the corresponding angular momentum loss. The amount of angular momentum lost with non-accreted material highly affects the orbital evolution of BH-BH system progenitors \citep{MacLeod2020,Willcox2023,Gallego0823,vanSon2022}.
The amount of specific angular momentum loss during non-conservative mass transfer is often implemented in population synthesis codes as a parameter $\gamma$. The below formula then gives the response of the orbital separation to binary mass transfer \citep{Tauris2006}: 

\begin{equation} \label{eq. orbital_evol}
\frac{\dot{\rm a}}{\rm a}=-2 \frac{\dot{M}_{\rm don}}{M_{\rm don}} \left(1 - \beta \frac{M_{\rm don}}{M_{\rm acc}} - (1-\beta) (\gamma + \frac{1}{2}) \frac{M_{\rm don}}{M} \right)
\end{equation} 

Where: $\beta$ - a fraction of accreted mass, $\gamma$ - dimensionless specific angular momentum of material ejected from binary. \\

The commonly adopted assumption is that the ejected material is lost with the specific angular momentum of the accretor $l_{acc} = \frac{M_{\rm don}}{M_{\rm tot}} \sqrt{GM_{\rm tot}a}$ which implies $\gamma_{\rm acc}=\frac{M_{\rm don}}{M_{\rm acc}}$. That is also the default value in our simulations for a BH accretor. Detailed studies using three-dimensional hydrodynamic models of binary coalescence indicate that the value of $\gamma$ depends on the system properties, such as mass ratio \citep{MacLeod2020}. In particular, the true amount of angular momentum loss might be significantly higher than $\gamma_{\rm acc}$, varying in the range between $\gamma_{\rm acc}$ and the specific angular momentum of the outer Lagrange point, $l_2 \approx 1.2^2 \sqrt{GM_{\rm tot}a}$ corresponding to $\gamma_{\rm L_{2}} \approx 1.2^2 {M_{\rm tot}}/{(M_{\rm don}M_{\rm acc})}$.

\begin{figure*}[!htbp]
\includegraphics[width=0.49\textwidth]{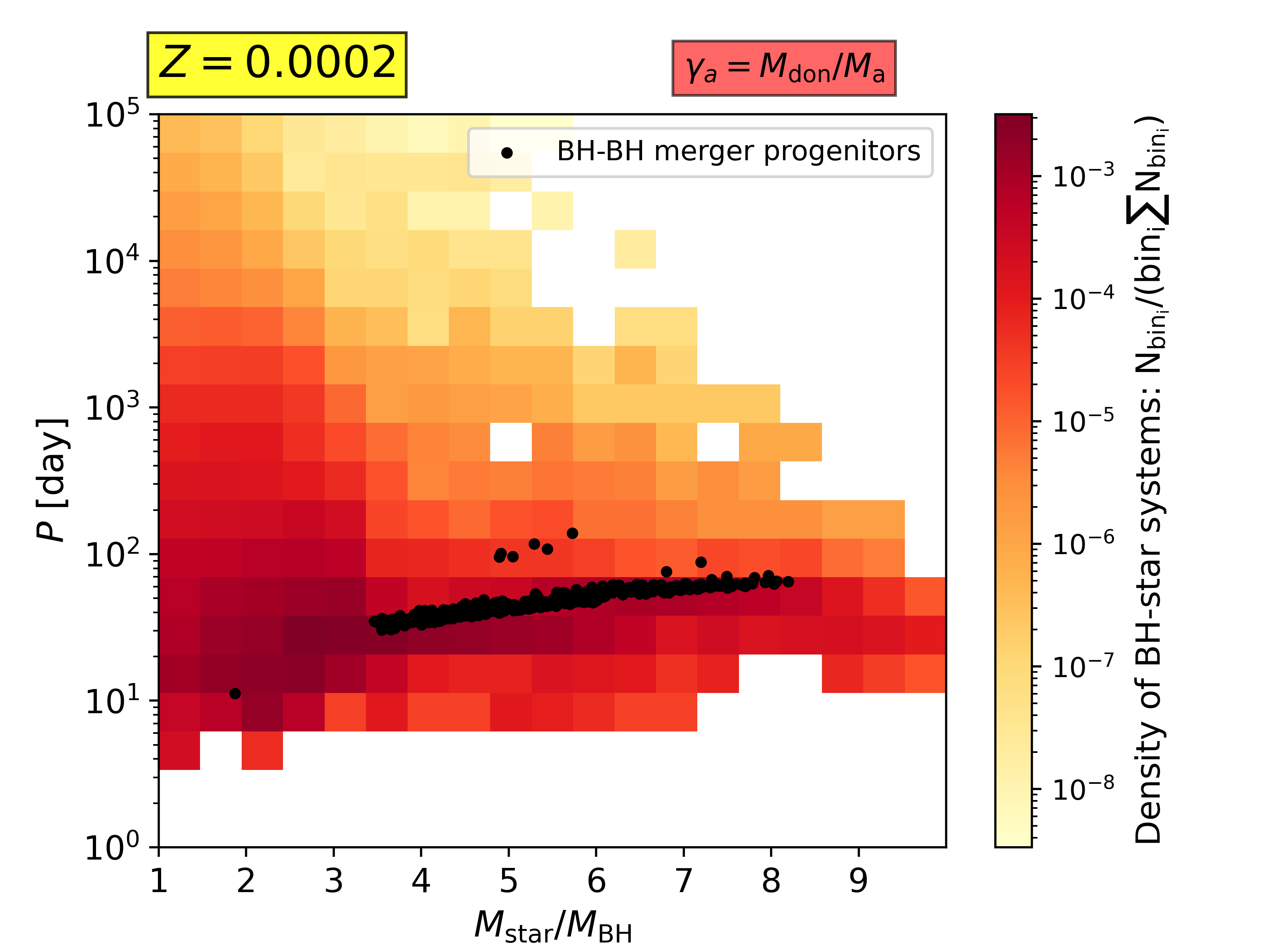}
\includegraphics[width=0.49\textwidth]{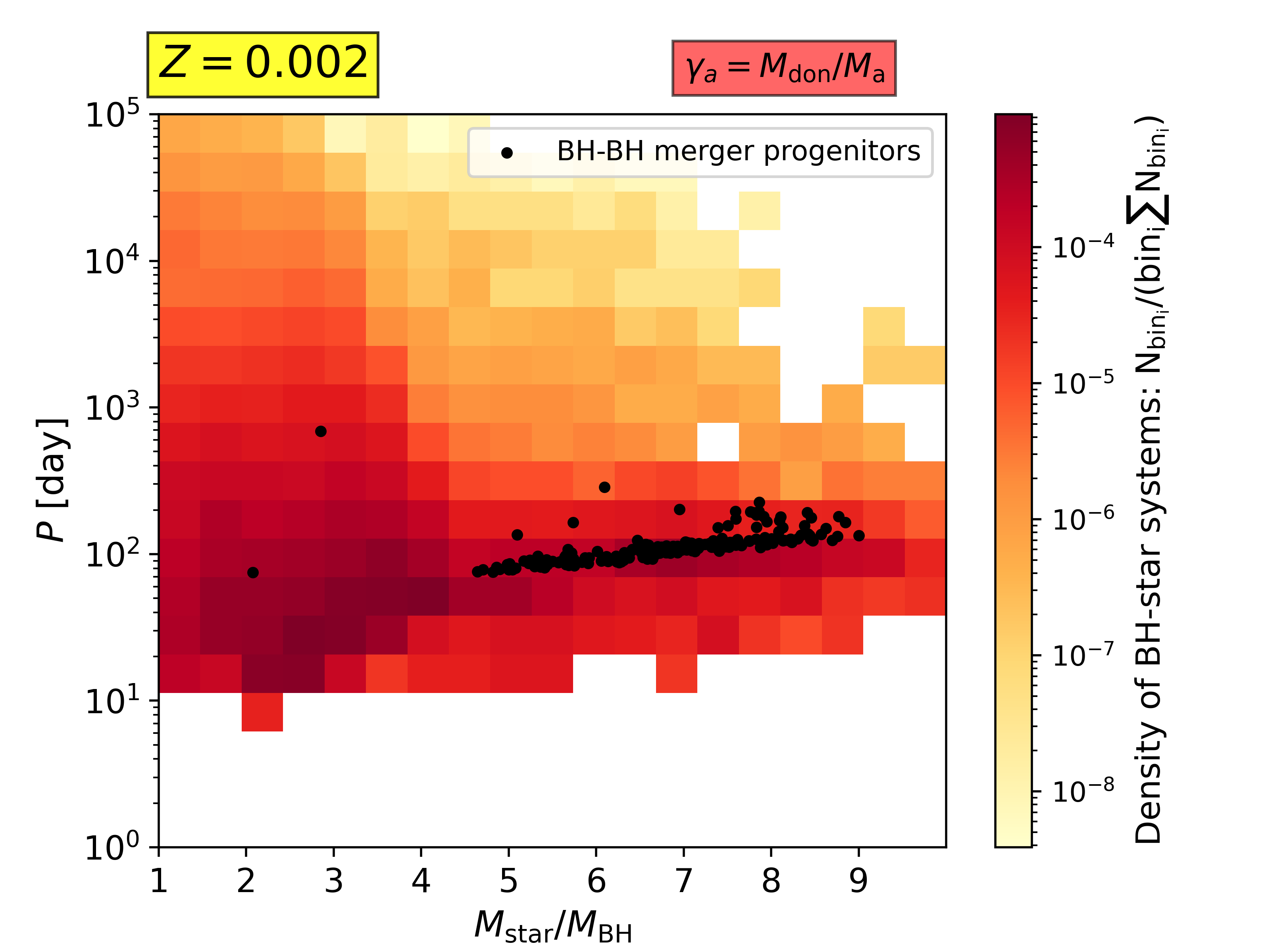}\\
\includegraphics[width=0.49\textwidth]{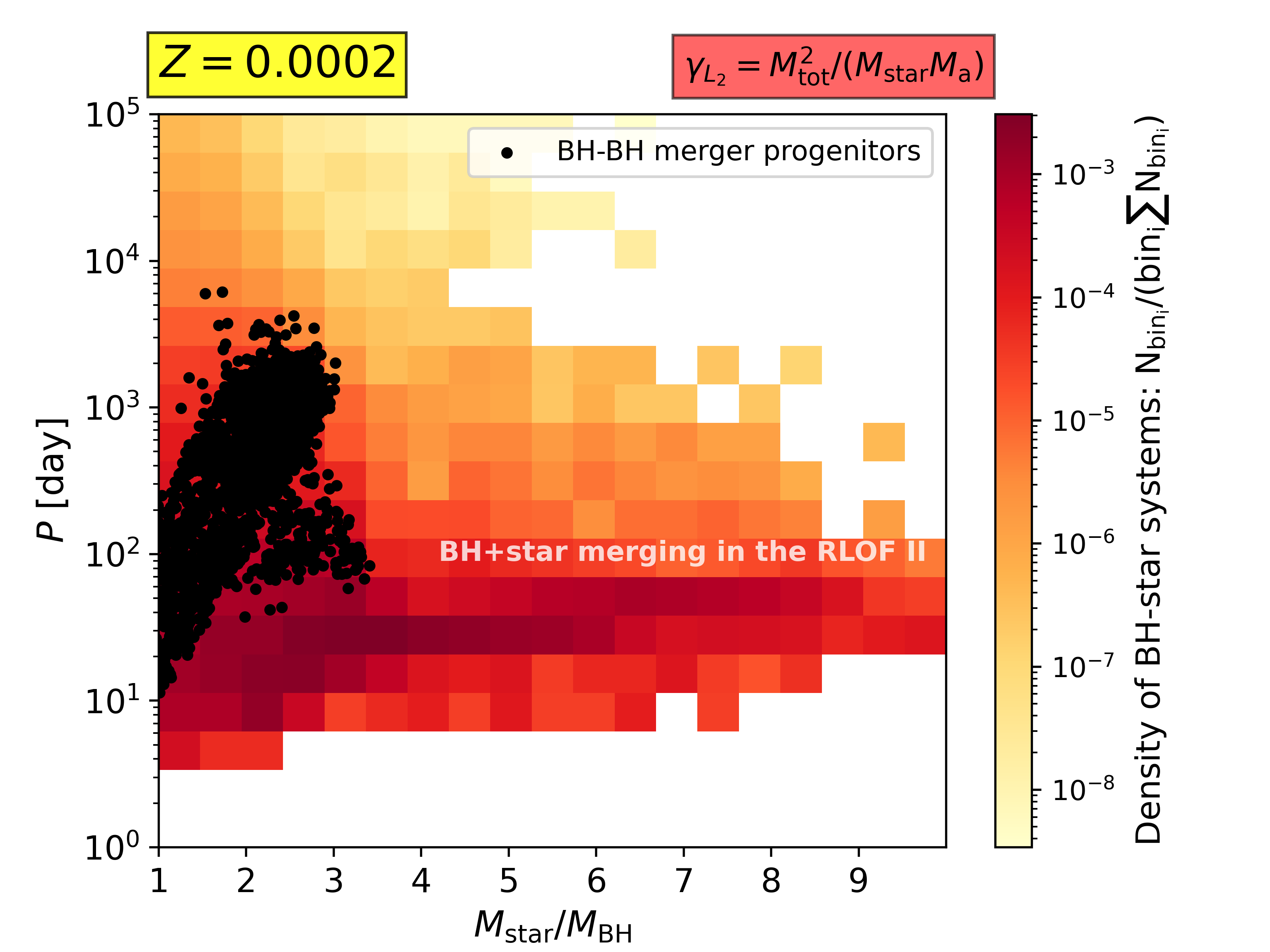}
\includegraphics[width=0.49\textwidth]{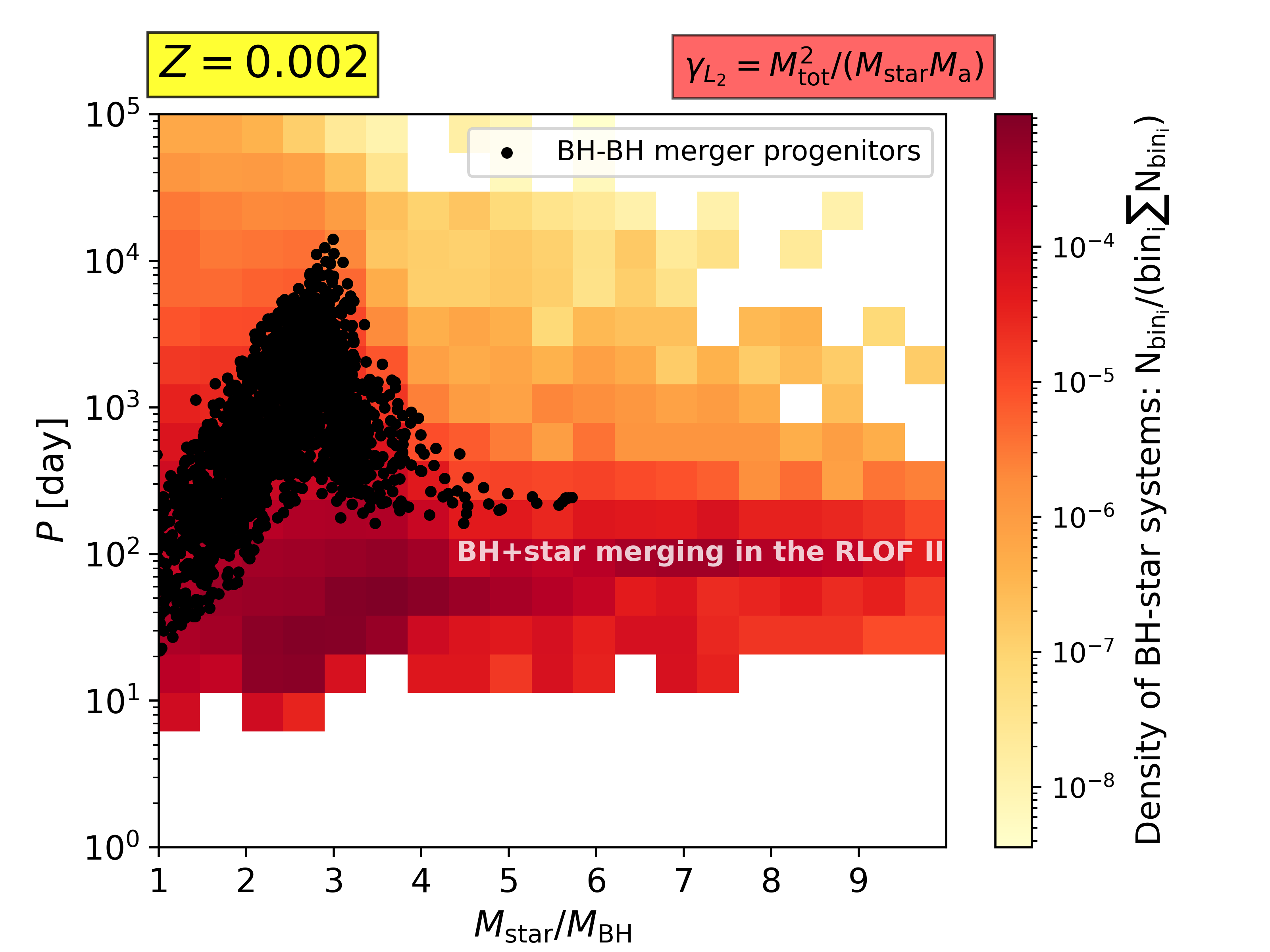}
\caption{Orbital period vs mass ratio distribution of BH binaries with a non-compact companion (mainly main sequence stars). Results only for the stable mass transfer subchannel (do not include the systems that evolved or would evolve through a common envelope). With the black dots, we mark progenitors of BH-BH mergers: on the top panel, the systems that would become BH-BH merger progenitors once adopting specific angular momentum loss of the accretor. On the bottom panels, black dots mark the systems that would become BH-BH progenitors under assumptions of high specific angular momentum loss, corresponding to the outer Lagrangian point. Left panels are for systems with $Z$ = 1\% $Z_{\odot}$, right panels with $Z$ = 10\% $Z_{\odot}$. Results for 100,000 generated binary systems with their initial masses in range $M_{1}>15 M_{\odot}$ and $M_{2}>$5 $M_{\odot}$.}
\label{fig: P_q_BH_MS}
\end{figure*} 

We test how the change in adopted angular momentum loss would impact the properties of BH-BH mergers. In particular, instead of $\gamma = \gamma_{\rm acc}$ we adopt a much higher value $\gamma = \gamma_{\rm L_{2}}$ accompanying accretion on a BH companion during the second mass transfer phase. We find that our synthetic population of BH-BH mergers is sensitive to our assumption on angular momentum loss. In particular, the tendency of stable mass transfer to produce distribution dominated by unequal mass ratio BH-BH mergers is no longer valid once we adopt the high value of $\gamma = \gamma_{\rm L_{2}}$. 

Figure \ref{fig: P_q_BH_MS} presents the orbital period (in days) vs mass ratio distribution of BH binaries with a non-compact companion at the moment of the first BH formation. The grids include all BH-star binaries in our simulations that evolved from 100,000 massive binary star systems with their initial masses $M_{1}>15 M_{\odot}$ and $M_{2}>$5 $M_{\odot}$. We show results for the two different metallicities: 1\% solar ($Z=0.0002$, left panels) and 10\% solar ($Z=0.002$, right panels). On top of the density plot, we marked by the black dots the systems that later evolved in close BH-BH binaries with their merger time below $t_{\rm GW}< 14$ Gyr. Note that Figure \ref{fig: P_q_BH_MS}, in contrast to Figure \ref{fig: Period_q_SMT_and_CE}, does not include the systems that at some point of their evolution went (or would go) through a CE phase. We selected only the systems evolving through stable mass transfer phases, both for underlying BH-star binaries (red and yellow density plots) and BH-BH merger progenitors marked as the black dots. As between the two tested models shown at the top and bottom panels, we only modified the angular momentum loss in the case of BH accretion, only the systems that later evolve into BH-BH mergers are affected (black dots). The underlying population of BH-star binaries for each metallicity remains the same.

The black dots in the top panels of Figure \ref{fig: P_q_BH_MS} show results for our default model with $\gamma = \gamma_{\rm acc}$ during the second, non-conservative mass transfer phase with a BH accretor. This model corresponds to relatively low angular momentum loss. In the bottom panels, we demonstrate how BH-BH progenitors are affected once we adopt high values of $\gamma = \gamma_{\rm L_{2}}$ instead, corresponding to loss of material with the specific angular momentum loss of the outer Lagrangian point. We find that our default model strongly favors the formation of BH-BH mergers from highly unequal-mass BH-star ($M_{\rm star}/M_{M_{\rm BH}}\geq 3$) systems and with moderate separations of $P \in 10-100$ days. BH-star systems with more equal components or with wide orbital separations evolve into wide BH-BH binaries instead (see more in Sec. \ref{sec: time_delays}). However, with increased angular momentum loss, progenitors of BH-BH mergers are much more diverse in terms of their mass ratios and period distribution. With high $\gamma = \gamma_{\rm L_{2}}$, wide BH-star binaries ($P \geq 100$ days) as well tight but more equal-mass systems ($M_{\rm star}/M_{M_{\rm BH}}\leq 2$) can form close BH-BH binaries that merge in 14 Gyr time. Increased angular momentum loss, on the other hand, makes BH-star binaries that are progenitors of BH-BH mergers in the default model, merge during the second mass transfer phase. Those systems, with highly unequal masses at the RLOF onset, rapidly shrink their orbits during the mass transfer to the moment when orbital separation no longer fits the donor radius. 

Narrow and characteristic parameter space for BH-BH progenitors in our default model is determined by the overlap of several factors. In particular: \\
\begin{enumerate}
\item Mass ratio reversal during the first RLOF results in a significant number of BH and massive-star binaries with unequal masses and with moderate orbital periods of $P \in 10-100$ days. Efficient accretion during the first mass transfer phase (typically case A or B) suppresses the formation of BH-star binaries with tight orbits with periods below 5–10 days.
\item Only unequal-mass BH-star binaries with a donor several times more massive than a BH might shrink during the stable mass transfer of their initial orbit of typically $P \in 10-100$ days to form a close BH-BH system that merges in 14 Gyr time.  
\item Unequal systems with wide orbital separations $P \geq 10-100$ go through the CE phase, whether they survive or not. Also, unequal, tight binaries might experience unstable mass transfer in our default model (see Appendix \ref{Sec: exp_instab}).
Note that the rapid population synthesis code could underestimate the mass of the donor's core (later the firstborn BH) after the first mass transfer in the case a star is still on its main sequence initiated it, see e.g. \cite{RomeroShaw2023}. That, besides the assumption on mass transfer efficiency, could lead to more unequal mass ratios and higher separations of BH-star systems than predicted by detailed stellar codes (see more in Sec. \ref{sec:discussion}).
\end{enumerate}

\begin{figure} 
\includegraphics[width=0.49\textwidth]{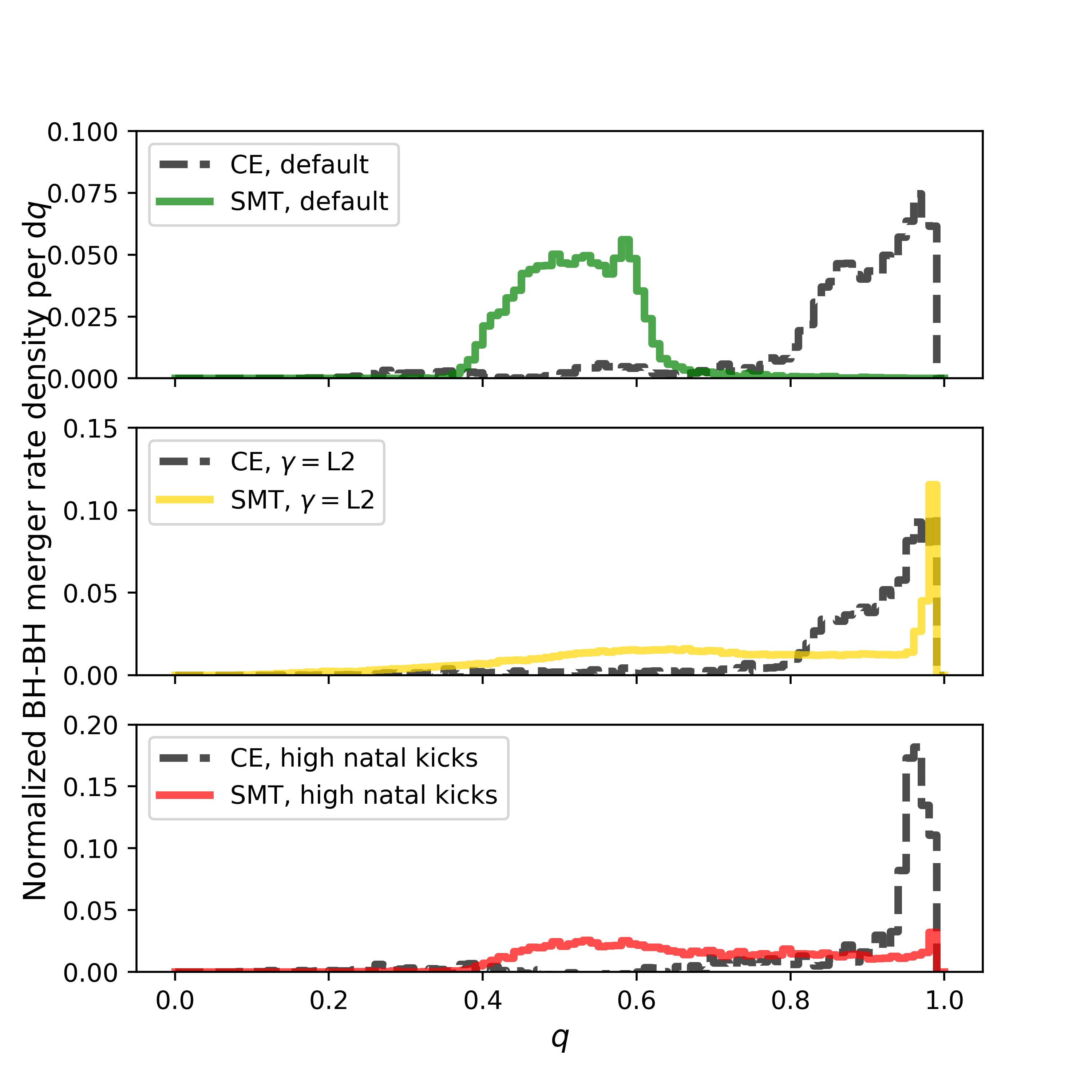}
\caption{Mass ratio $q$ distributions of BH-BH mergers (at redshifts $z<2.0$) formed via stable mass transfer formation scenario (SMT, colorful, continuous line) and common envelope scenario (CE, dashed line) separately for three tested models. The total sum normalizes the merger rate densities of BH-BH systems. Results for our default model (upper panel), a model with high angular momentum loss $\gamma = \gamma_{\rm L_{2}}$ (middle panel), and a model with non-fallback decreased (full) natal kicks (bottom panel), see Section \ref{sec:method}.}
\label{fig: mass_ratio_three_models}
\end{figure} 

\subsection{Time delays} \label{sec: time_delays}

The orbits of BH-BH progenitors are affected in different ways by unstable and stable mass transfer due to their various time scales, conservation of mass, and accompanying angular momentum loss. Therefore, the final properties of BH-BH merger subpopulations produced via those two subchannels might significantly differ. Due to our still limited understanding of those processes, simplified models that rely on various uncertain assumptions are used to predict the final evolutionary outcome for the systems. The various models result in contrasting properties of formed GW sources, even within the same evolutionary scenario. Specific assumptions may leave characteristic fingerprints in the parameter distribution of formed BH-BH mergers. For example, we find that once adopting low amounts of angular momentum loss during the non-conservative mass transfer phase, stable mass transfer subchannel strongly favors the formation of unequal-mass BH-BH mergers \citep{Olejak2021a,Olejak2022}. In particular, it results in a characteristic, broad peak in the mass ratio distribution of BH-BH mergers between $q \in (0.4-0.7)$, see Figure \ref{fig: mass_ratio_three_models}. The reasons, why the peak emergence (e.g. due to orbital response to mass transfer) are described in Section \ref{Sec: ang_mom_loss}.

\begin{figure*}[!htbp]
\includegraphics[width=0.49\textwidth]{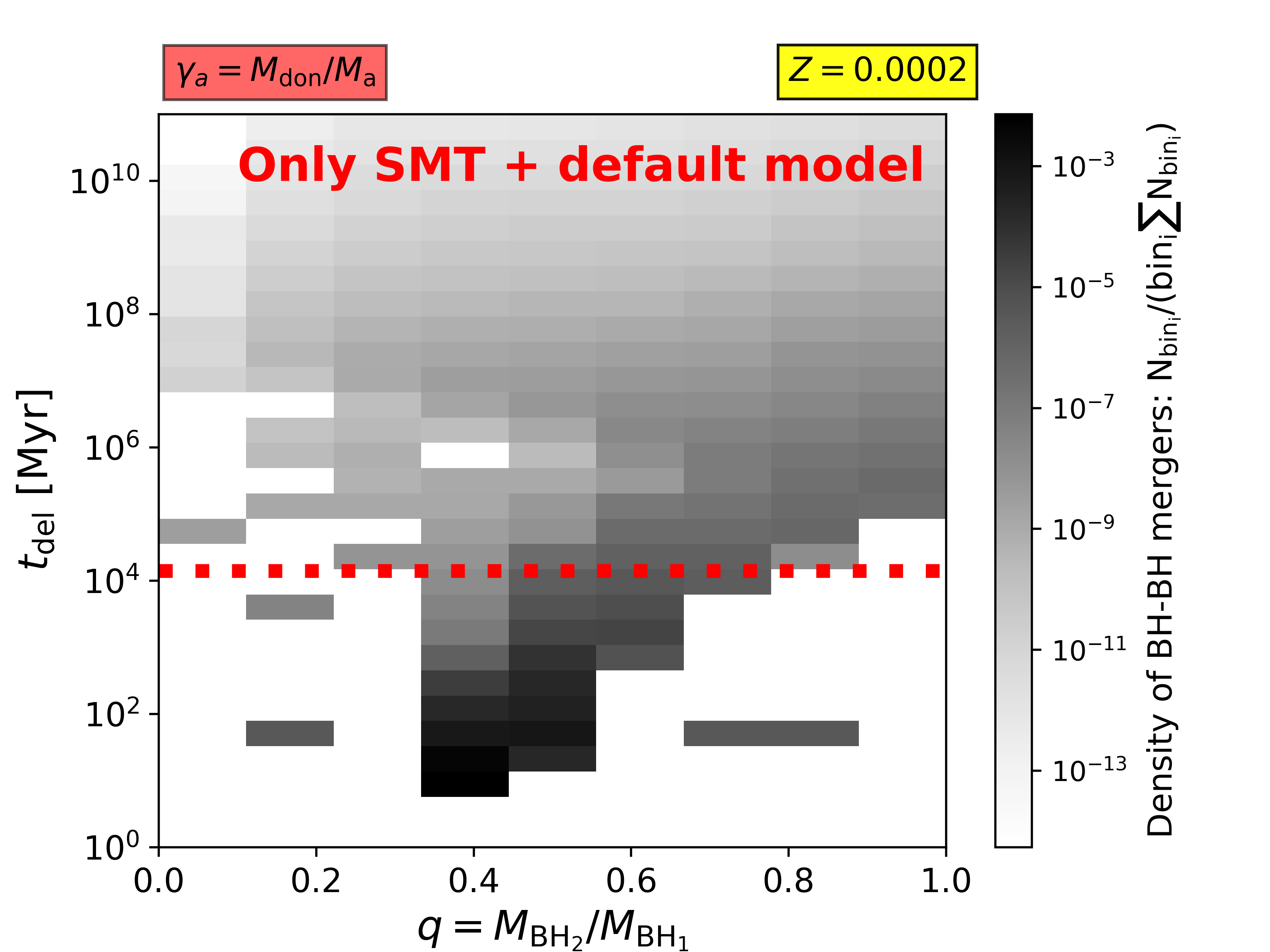}
\includegraphics[width=0.49\textwidth]{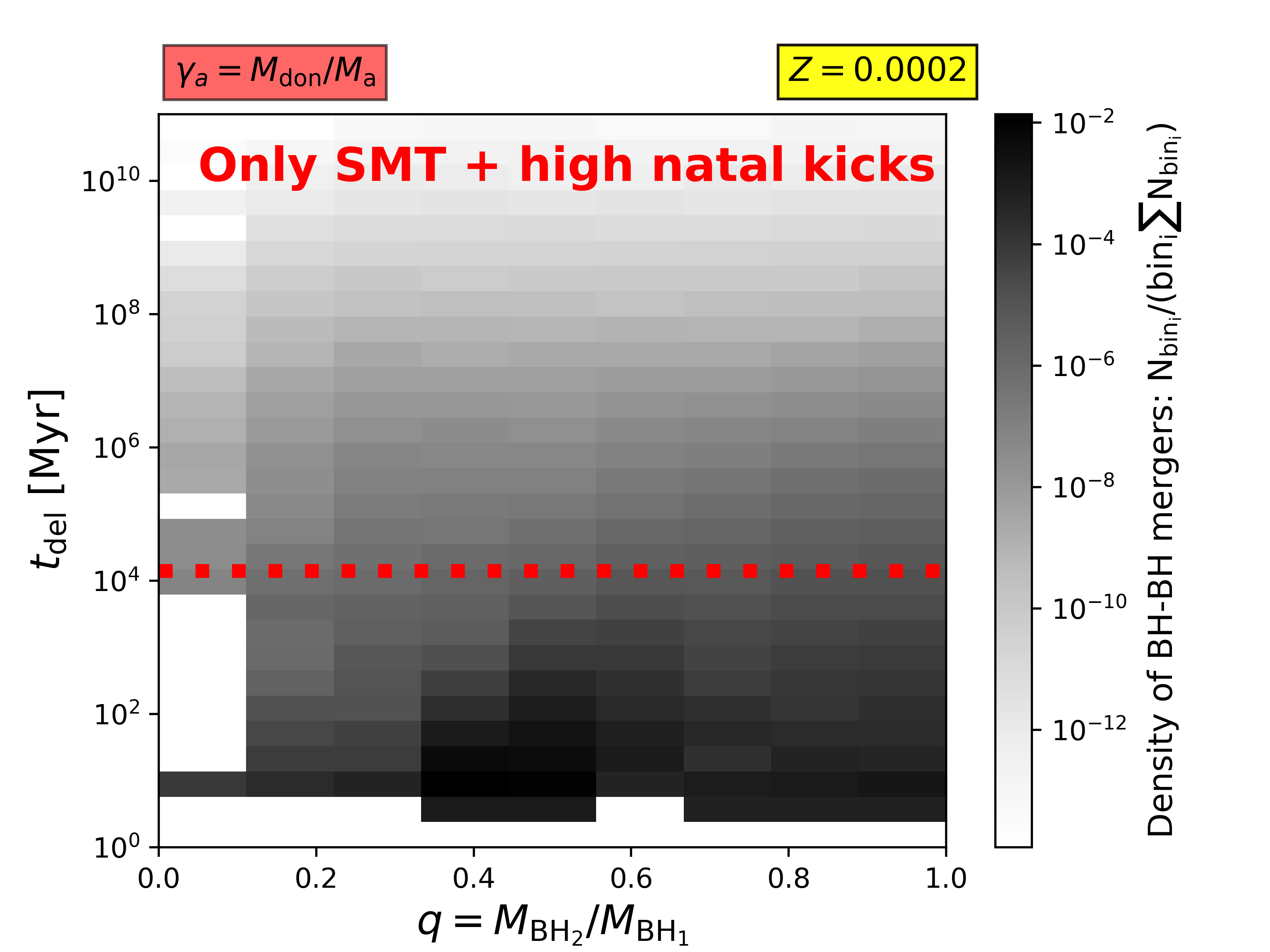}\\
\includegraphics[width=0.49\textwidth]{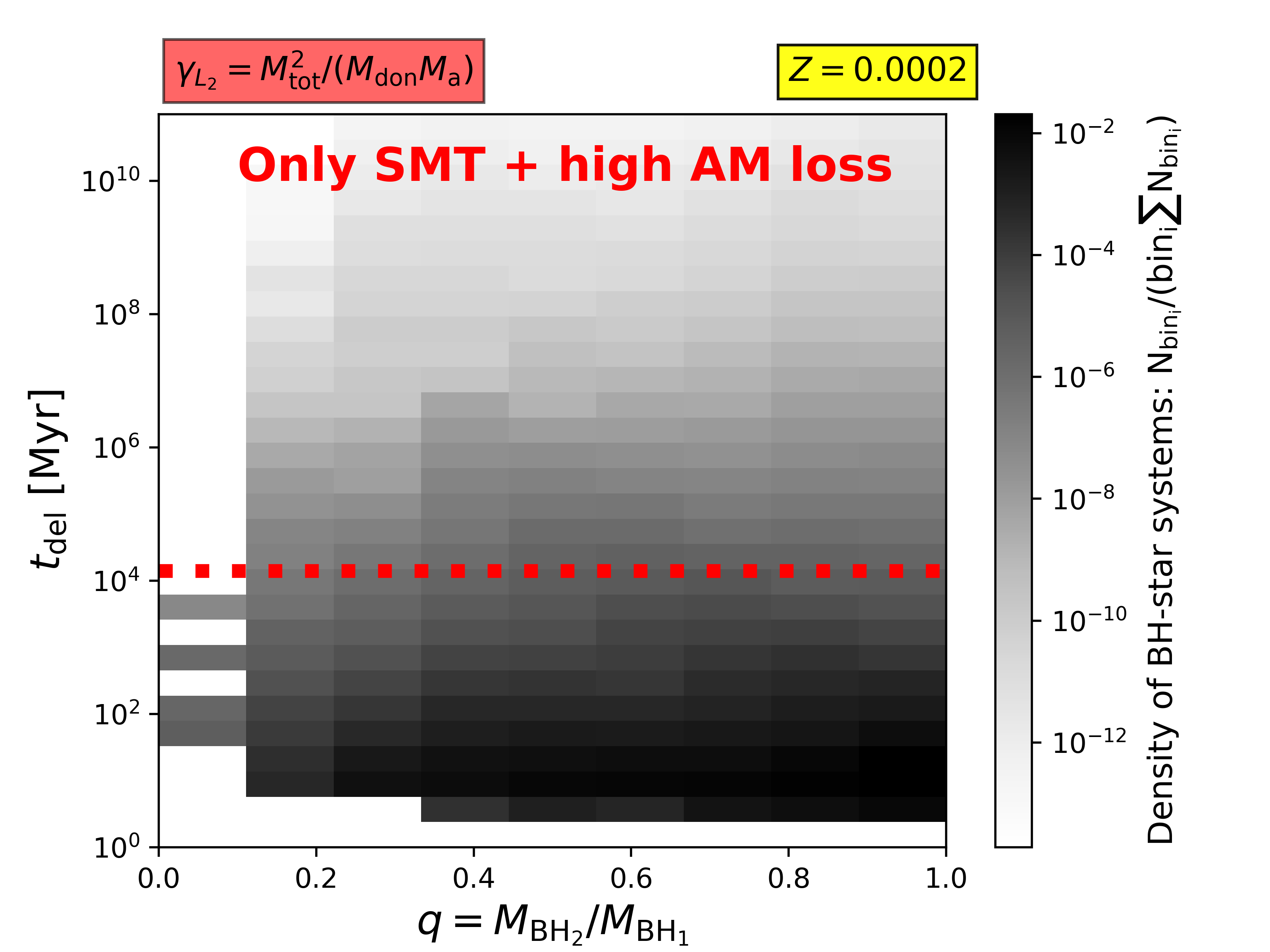}
\includegraphics[width=0.49\textwidth]{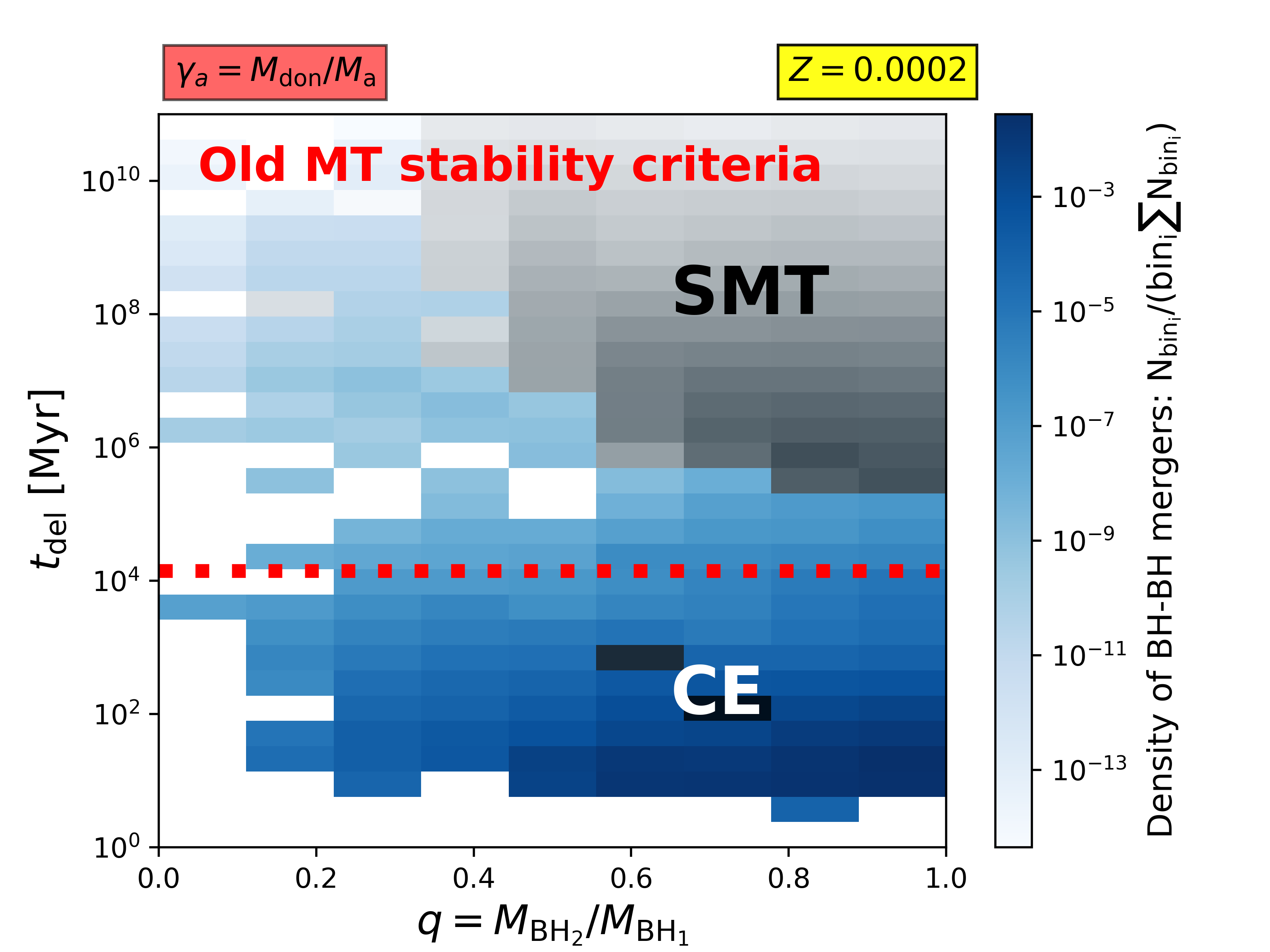}\\
\caption{Time delay ($t_{\rm del}$[Myr]) vs mass ratio distribution of BH-BH systems in four tested models. The red, dashed line corresponds to 14 Gyr. Top, left panel: BH-BH mergers formed via stable mass transfer subchannel in our default model with revised mass transfer stability criteria. This model includes fallback, decreased natal kicks, and low angular momentum loss during non-conservative mass transfer. Top, right panel: BH-BH binaries formed via stable mass transfer subchannel in the model with revised mass transfer stability criteria and low angular momentum loss but non-fallback decreased natal kicks. Bottom, left panel: BH-BH systems formed via stable mass transfer subchannel in the model with revised mass transfer stability criteria and fallback decreased natal kicks (as in our default model) but high angular momentum loss during non-conservative mass transfer ($\gamma=\gamma_{\rm L_{2}}$). Bottom, right panel: BH-BH mergers formed by both the CE with $\alpha=1.0$ (dominant) and the stable mass transfer (subdominant) subchannels in the model with old criteria for mass transfer stability.}
\label{fig: Time_delays}
\end{figure*}

The preference towards unequal-mass BH-BH mergers in stable mass transfer subchannel is well illustrated by the distribution of the BH-BH systems' time delays as a function of their mass ratios, see Figure \ref{fig: Time_delays}. The red dashed line in the panels separates close BH-BH binaries that merge within 14 Gyr (below) from wide BH-BH systems with time delays above 14 Gyr (up). The left, upper panel of the Figure shows the results for our default model with low angular momentum loss $\gamma = \gamma_{\rm acc}$. The subpopulation of BH-BH systems with short time delays $t_{\rm del}<14$Gyr consists mainly of the systems with mass ratios in the range of $q \in 0.4-0.7$. Importantly, the same model produces numerous BH-BH systems with nearly equal masses $q \approx 1.0$. However, their orbits are much too wide to merge within the time of the Universe age. The typical time delays of the equal-mass BH-BH systems vary in the range of $t_{\rm del} \in 10^{6} - 10^{11}$ Gyr.

The relative fraction of equal to unequal-mass BH-BH systems that merge within 14 Gyr is increased once adopting different prescriptions for the natal kicks - see the right, top panel of Figure \ref{fig: Time_delays}. Once applying high, non-fallback decreased BH natal kicks we significantly decrease the total number of formed BH-BH systems. However, high kicks allow relatively more systems with comparable component masses to merge in a short time due to their eccentric orbits. Note that time delay of the BH-BH system merger  $t_{\rm del} \sim (1-e^2)^{7/2}$ \citep{1964PhDT........51P}. See more in subsection \ref{sec:natal_kicks} devoted to the impact of natal kicks and core collapse physics. 

The bottom, left panel of Figure \ref{fig: Time_delays} demonstrates that BH-BH mergers mass ratio distribution in our models is highly sensitive to the adopted assumption on the specific angular momentum loss. In this model, the non-accreted material during the second RLOF is lost from the system with the specific angular momentum of the outer Lagrangian point. Increased angular momentum loss leads to more efficient orbital contraction during the highly non-conservative mass transfer on a BH accretor than in our default model. Therefore, the progenitors of equal-mass component BH-BH systems, which in our default model with low $\gamma = \gamma_{\rm acc}$ tend to finish evolution with wide orbits (time delays orders of magnitudes above the age of the Universe), now can merge in the time of 14 Gyr. As the highly unequal-mass BH-star systems merge during the ongoing mass transfer phase accompanied by high angular momentum loss, in this model, equal-mass component systems are common among BH-BH mergers. 

Finally, in the bottom, right panel of Figure \ref{fig: Time_delays} we show results for the model in which BH-BH mergers, in contrast to our default model, are formed mainly via the CE subchannel \citep{Olejak2021a}. This model is much looser in terms of CE development than our default model with revised mass transfer stability criteria. BH-BH systems in this model are characterized by rather short time delays, typically below a few Gyr. Distribution is dominated by equal-mass BH-BH mergers. Progenitors in the CE channel, contrary to the stable mass transfer scenario, do not experience such a significant mass-ratio reversal during the first RLOF. The initial, zero-age main sequence, separation of BH-BH mergers progenitors in the CE subchannel is usually 1-2 orders of magnitude larger than stable mass transfer progenitors, so the first mass transfer is usually initiated at the later part of evolution, once the donor has already well-defined core-envelope boundary.  

Our results indicate that the distribution of GW sources dominated by equal-mass BH-BH mergers is consistent with both stable mass transfer formation scenarios accompanied by efficient angular momentum loss, and the parametrized common envelope prescription with $\alpha = 1.0$. The inference of GW detections implying a high fraction of unequal-mass BH-BH mergers with their mass ratio $q \in (0.4-0.7)$, could indicate the contribution of a stable mass transfer scenario with non-enhanced angular momentum loss. The mass ratio distribution of BH-BH mergers in the CE subchannel, however, can be also affected by core-collapse physics or mass transfer efficiency, see e.g. \cite{Olejak2022}.

\subsection{Evolutionary scenario} \label{sec: evolutionary_scenario}

\begin{figure} 
\includegraphics[width=0.49\textwidth]{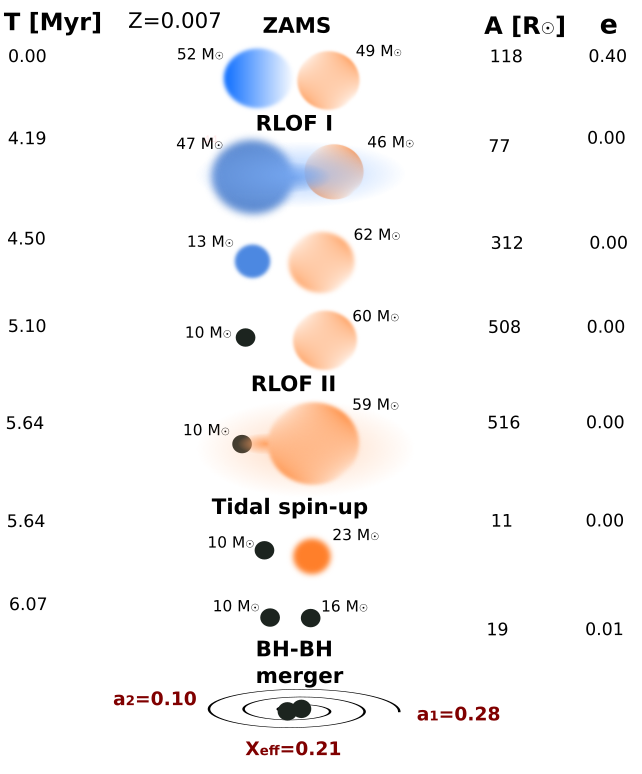}\\
\caption{Typical stable mass transfer scenario for the formation of BH-BH merger with unequal component masses and high, positive effective spin $\chi_{\rm eff}>0.1$. }
\label{fig: evol_scen}
\end{figure}

In this subsection, we present and describe the typical evolutionary scenario for the formation of highly spinning, unequal-mass BH-BH mergers via the stable mass transfer subchannel in our default model. Figure \ref{fig: evol_scen} shows the main stages of the evolution of a binary system.

In contrast to the classic CE scenario (see evolutionary diagrams in e.g. \cite{Belczynski2016b,Olejak2021b}), BH-BH mergers progenitors in our stable mass transfer subchannel typically start their evolution on relatively tight orbits with their initial separation of several dozen to hundreds of solar radii. This makes binaries go through the first RLOF phase while the donor is usually shortly after the main sequence, in the early phase of rapid expansion. The important point of this evolutionary scenario is mass ratio reversal after the first stable mass transfer phase. The companion (initially less massive) star gains a large fraction of material transferred from the donor, significantly increasing its mass. The donor loses most of its initial mass and gets stripped of its hydrogen envelope, remaining just a helium core.

At the moment of the first BH formation, the mass ratio of the binary components is highly unequal (donor typically 3-7 times more massive than its BH companion) and it has a moderate orbit of $P \in 10-100$ days, see Figure \ref{fig: P_q_BH_MS}. Then, our BH-BH progenitors go through the second stable mass transfer phase. The donor is usually an early-type, radiative envelope giant, rapidly expanding after leaving its main sequence. The unequal mass ratio at the onset of the second RLOF allows systems to effectively shrink their orbits in highly non-conservative mass transfer on a BH accretor. The separation and mass ratio systems at the RLOF onset may be so tuned to the loss of angular momentum, that the donor is stripped of its hydrogen envelope when the mass transfer still leads to orbital contraction, producing a very tight BH-helium core binary. Note that systems with such unequal mass ratios but with wider orbits initiate an unstable mass transfer phase within our revised mass transfer stability criteria. 
If after the second RLOF, the orbital period of the BH-helium core binary is below $P_{\rm orb} \lessapprox 1$ day it enters the regime of efficient tidal spin-up of the core by a BH. Such an evolutionary scenario (already suggested and described by \cite{Olejak2021b} and \cite{Broekgaarden2022}) leads to the production of the BH-BH binary in which the second-born high-spinning BH is also the more massive one.

\subsection{The role of natal kicks} \label{sec:natal_kicks} 

\begin{figure*} 
\includegraphics[width=0.49\textwidth]{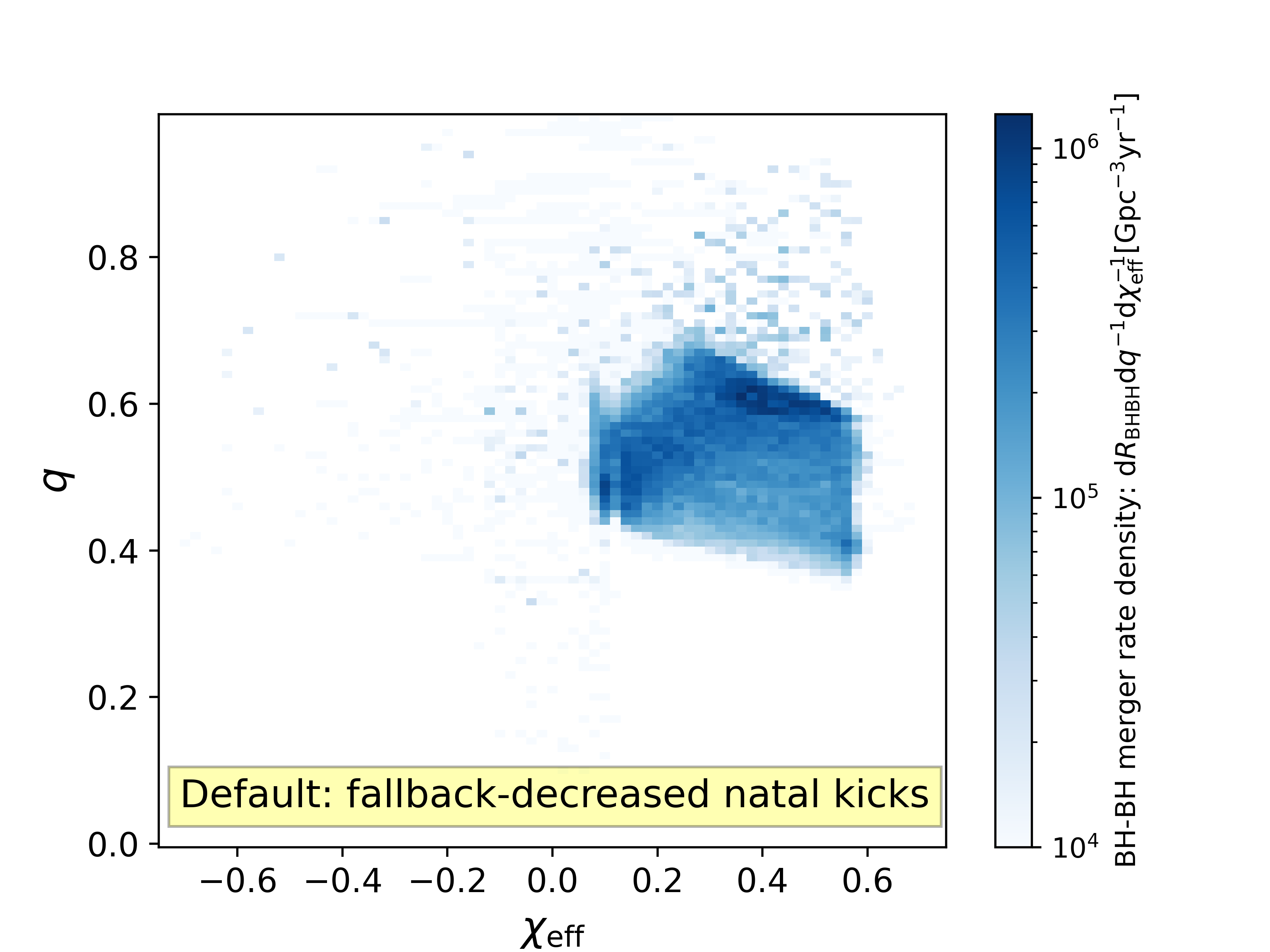}
\includegraphics[width=0.49\textwidth]{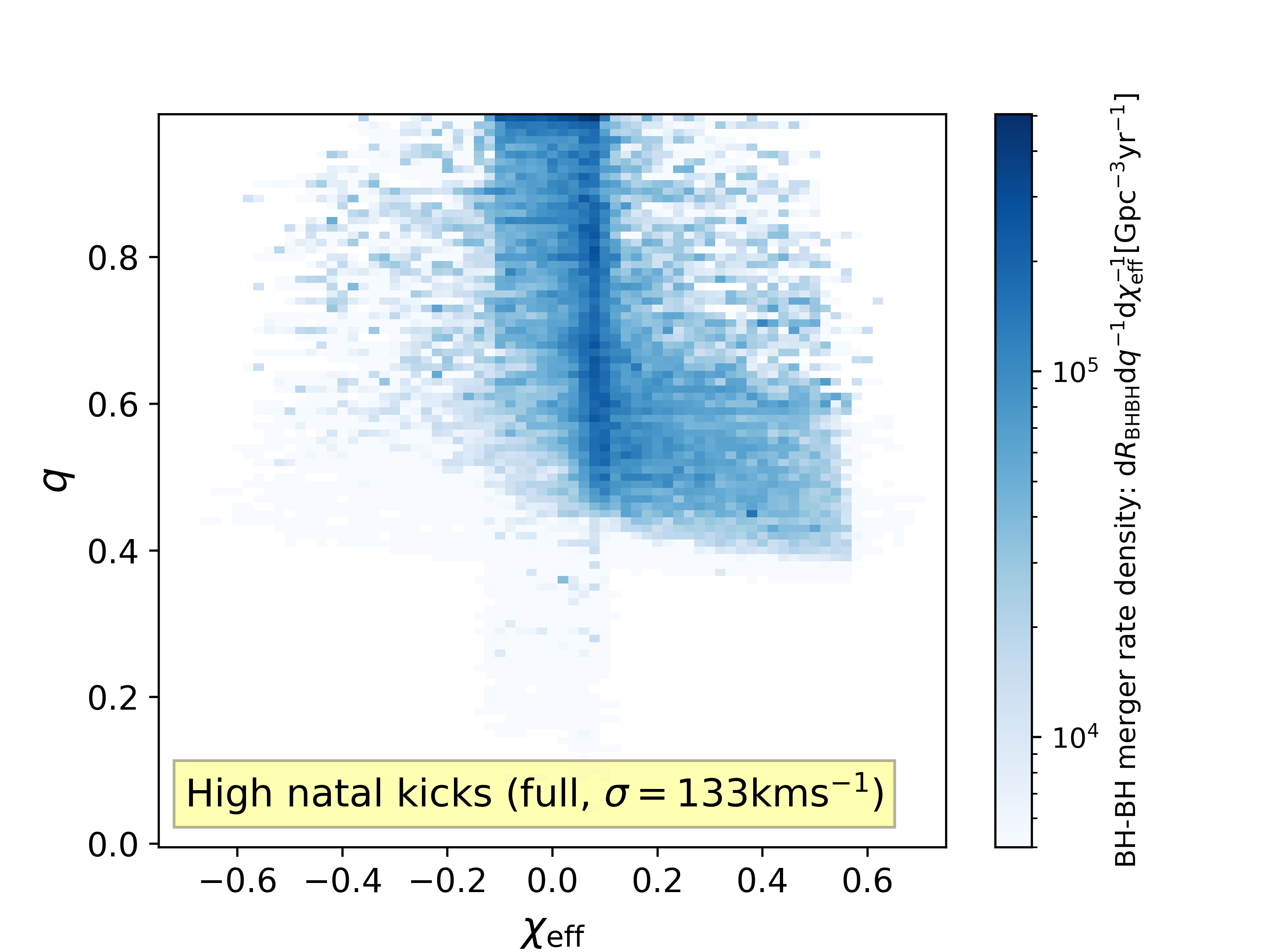}
\caption{Two-dimensional $\chi_{\rm eff}$ - mass ratio $q$ histograms of BH-BH merger rates density $\frac{\rm dR_{\rm BHBH}}{{\rm dq}{\rm d} \chi_{\rm eff}}$[Gpc$^{-3}$yr$^{-1}$] at redshift $z<1.0$. Only the systems formed via stable mass transfer subchannel (CE subchannel not included). Results for two natal kick variants: on the left panel a model with fallback-decreased natal kicks (default) and on the right panel with non-fallback decreased magnitudes (full kick).}
\label{fig: natal_kicks}
\end{figure*}

The mechanism of the natal kicks accompanying compact object formation is uncertain. Especially, in the case of BHs, the available constraints are very weak. It is expected that the formation of BHs could be followed by partial or even full fallback of matter on the proto-NS \citep{Fryer2012}. The magnitudes of natal kicks are usually expected to be significantly lower for BHs than for neutron stars \cite{Janka2024}. In rapid population synthesis codes such as {\tt StarTrack}, this relation is approximated by decreasing the possible magnitude of the kick inversely proportional to the amount of fallback on the newborn compact object \citep{Fryer2012, Belczynski2012}. Then, progenitors of massive BHs with their pre-SN cores above the mass threshold, finish their evolution in a "direct collapse", accompanied only by anisotropic emission of neutrinos. The final fate of the star might, however, depend on the details of its final pre-SN structure, which are not included in rapid population synthesis modeling
\citep{Sukhbold2014, Patton2020, Laplace2021, Schneider2021, Fryer2022, Janka2024}. In this subsection, we test how the properties of our synthetic BH-BH merger population are affected by various prescriptions for the final BH masses and the natal kicks they get.

In Figure \ref{fig: natal_kicks} we present 2-dimensional effective spins $\chi_{\rm eff}$ vs mass ratios $q$ histograms of BH-BH merger rate densities up to redshifts $z<1.0$. The histograms are only for BH-BH mergers formed via the stable mass transfer subchannel and for the two different BH natal kick models. On the left panel, we show the results for our default model with fallback-decreased natal kicks. On the right panel, we present results for the alternative approach with the high BH natal kicks (full, independent of the BH progenitor mass, see Sec. \ref{sec:method}). This approach is motivated by a lack of robust observational constraints and a rather poor theoretical understanding of the core collapse and the mechanism of BH natal kicks. High BH natal kicks so far cannot be ruled out but they could have interesting and important consequences for the $\chi_{\rm eff}$ distribution of merging BH-BH systems. In particular, they significantly increase the fraction of negative effective spin BH-BH mergers and the contribution of equal-mass events in our stable mass transfer channel. Note, however, that this model should be considered rather as an upper limit on the misaligned BH-BH mergers with $\chi_{\rm eff}$ and some recent studies suggest much lower BH kicks \citep[see e.g.][]{Janka2024}.

\begin{figure}
\includegraphics[width=0.48\textwidth]{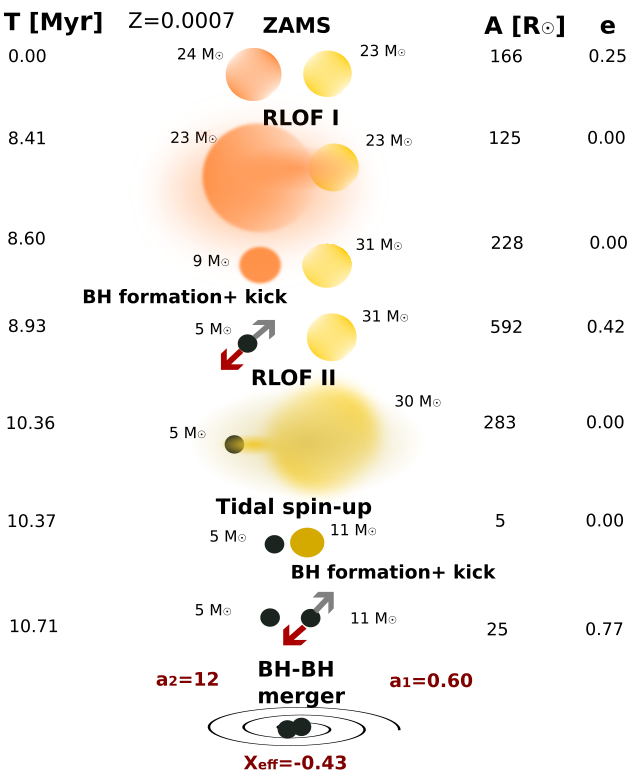}
\caption{Stable mass transfer formation scenario of BH-BH merger with unequal component masses and high, negative effective spin $\chi_{\rm eff} \approx -0.4$ in the model with full (non-fallback decreased) natal kick magnitudes. This model results in a higher fraction of negative $\chi_{\rm eff}<0$, see right panel of Figure \ref{fig: natal_kicks}. }
\label{fig: neg_eff_spin}
\end{figure}  

High BH natal kicks lead to more frequent disruption of the BH-BH system and a decrease in the total number of formed BH-BH mergers. However, if the BH-BH systems remain bounded, they possibly end evolution on highly eccentric orbits ($e >0.5$). High eccentricities, on the other hand, significantly decrease expected inspiral time as $T_{\rm ins} \sim (1-e^2)^{7/2}$ \citep[][]{1964PhDT........51P} allowing eccentric BH-BH systems on wide orbits to merge in a much shorter time (below 14 Gyr). High kicks also increase the fraction of BH-BH mergers with negative effective spins $\chi_{\rm eff}<0.0$. Figure \ref{fig: neg_eff_spin} presents an evolutionary diagram for the progenitor system of BH-BH merger that ends evolution with very high, negative effective spin value of $\chi_{\rm eff} \approx -0.4$. The formation scenario is similar to the one for the BH-BH merger with highly positive effective spin, presented in Figure \ref{fig: evol_scen}. However, in this case, after the efficient tidal spin-up of the helium core by the primary BH, the second-born BH obtains a high natal kick. That causes significant individual BH spin misalignment with the system's orbital momentum axes. The high eccentricity of the formed wide BH-BH system ($e \approx 0.8$) allows it to merge in the Hubble time.

\section{Discussion} \label{sec:discussion}

\subsection{Comparison with detailed stellar model grid}

In this section, we make a brief comparison of the BH-star binaries grid generated with our {\tt StarTrack} models with the grid obtained using detailed stellar evolution code. For that purpose, we adopt the BH-MS binaries predicted by \citep{Xuinprep}, which are derived from a dense grid of detailed binary evolution models by \cite{Wang2020}. This model grid is computed by the MESA code \citep[version 8845][]{Paxton2011,Paxton2013,Paxton2015} with the metallicity of the Small Magellanic Cloud, $Z \approx 0.002 \approx 10\% Z_{\odot}$. In the adopted detailed models, critically rotating stars can not accrete material, which results in near-zero mass transfer efficiencies for all Case B mass transfer. For case A mass transfer, the accretion efficiency is up to about 60\%. The star is assumed to end its evolution as a BH once a core helium-depleted star has its helium core mass above 6.6$\,\msun$, the outcome compact object is assumed to be a BH \citep{Sukhbold2018}. The final BH mass is computed by ejecting 20\% of the mass of the helium envelope and then losing 20\% of the mass of the remaining object due to the release of gravitational binding energy \citep{Kruckow2018}. Detailed grids do not include natal kicks for newborn BHs.

There are a few important differences between the input assumptions in our default {\tt StarTrack} model and the detailed model grid. The major ones are:\\
\begin{enumerate}
\item A high efficiency of mass transfer on non-degenerate accretor. In contrast to the detailed models, our default model allows for a much higher fraction ($\beta = 50\%$) of the mass accreted during the first RLOF. Donors in our stable mass transfer channel are usually early-type giants, at the beginning or in the middle of their expansion and for that type of binaries, mass transfer in our simulations is expected to be rather efficient. \footnote{Note that the true mass transfer efficiency is unknown, and both assumptions are justified by the lack of robust constraints. Also, previous studies trying to fit individual observations to determine the accretion fraction are rather divergent in their results \citep{Nelson2001,Mennickent2013,deMink2007}. Some recent studies even suggest that higher accretion efficiency during Case B than during Case A mass transfer would better reproduce observed systems \citep{RomeroShaw2023}.}
\item Natal kicks and core-collapse physics. Our default model, in contrast to detailed grids, considers the possible effect of BH natal kicks. This could affect the orbits of the BH-star systems, especially by reducing the number of tight systems with a low-mass BH. The choice of the core-collapse physics also determines the final BH mass. That impacts the mass ratio distribution of systems.
\item Possible underestimation of the post-RLOF donor cores in the case when the mass transfer was initiated during the main sequence. This is a common feature in rapid population synthesis codes which are based on stellar evolutionary tacks by \cite{Hurley_2002}. The stellar core is not modeled during the main sequence and when it initiates mass transfer, its mass is approximated by its fit to the stellar mass at the terminal-age main sequence. The underestimation of the donor's core mass during case A mass transfer might shift our mass ratio distribution towards more unequal and wider binaries. Some recent rapid codes implemented ad-hock methods to minimize this effect, see \cite{RomeroShaw2023}.
\end{enumerate}
All of these differences tend to shift BH-star binaries produced by the default {\tt StarTrack} model towards the more unequal mass ratios and wider orbital separations than the binaries predicted by the detailed grid. 

\begin{figure*} 
\includegraphics[width=0.5\textwidth]{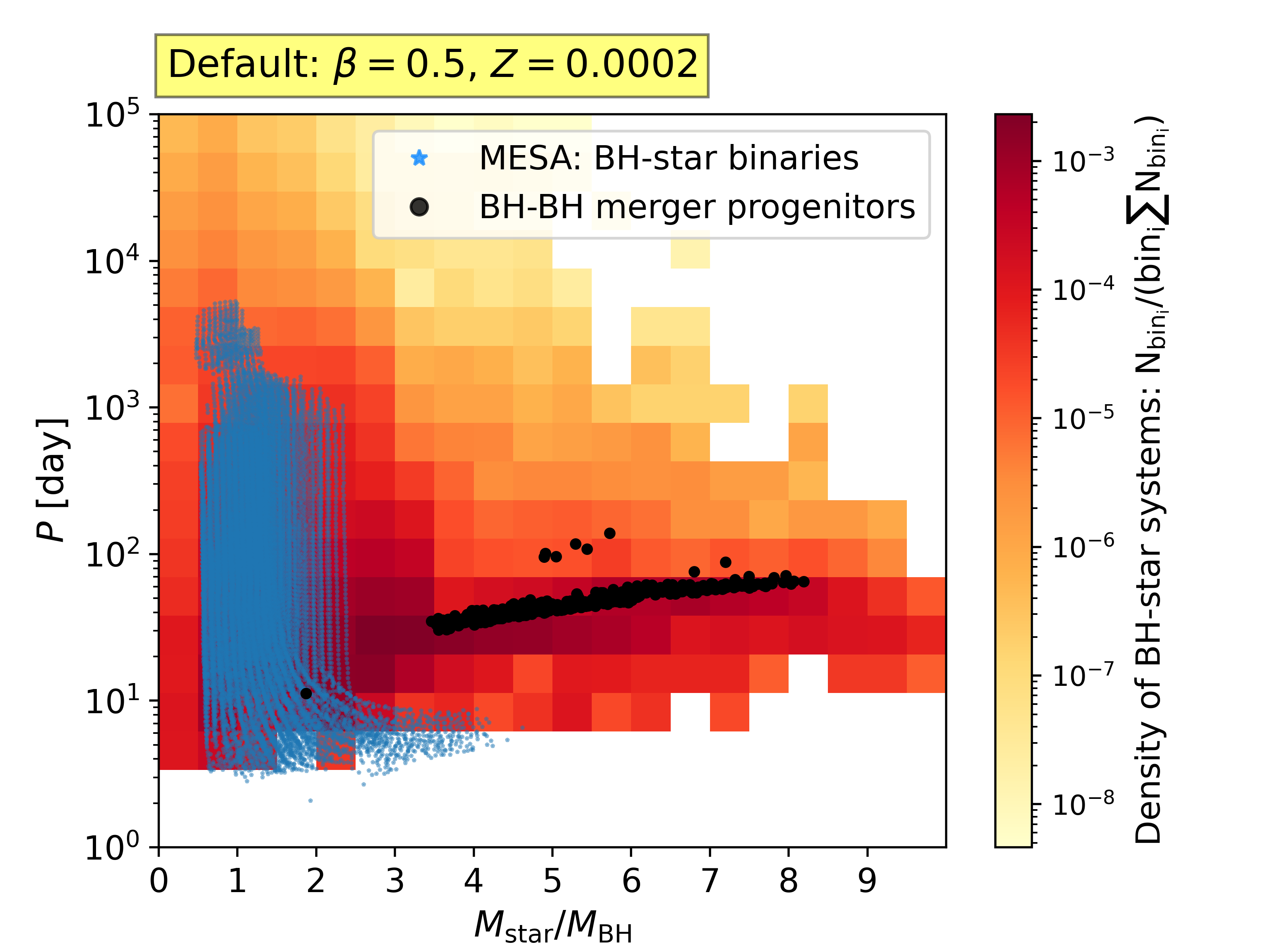}
\includegraphics[width=0.5\textwidth]{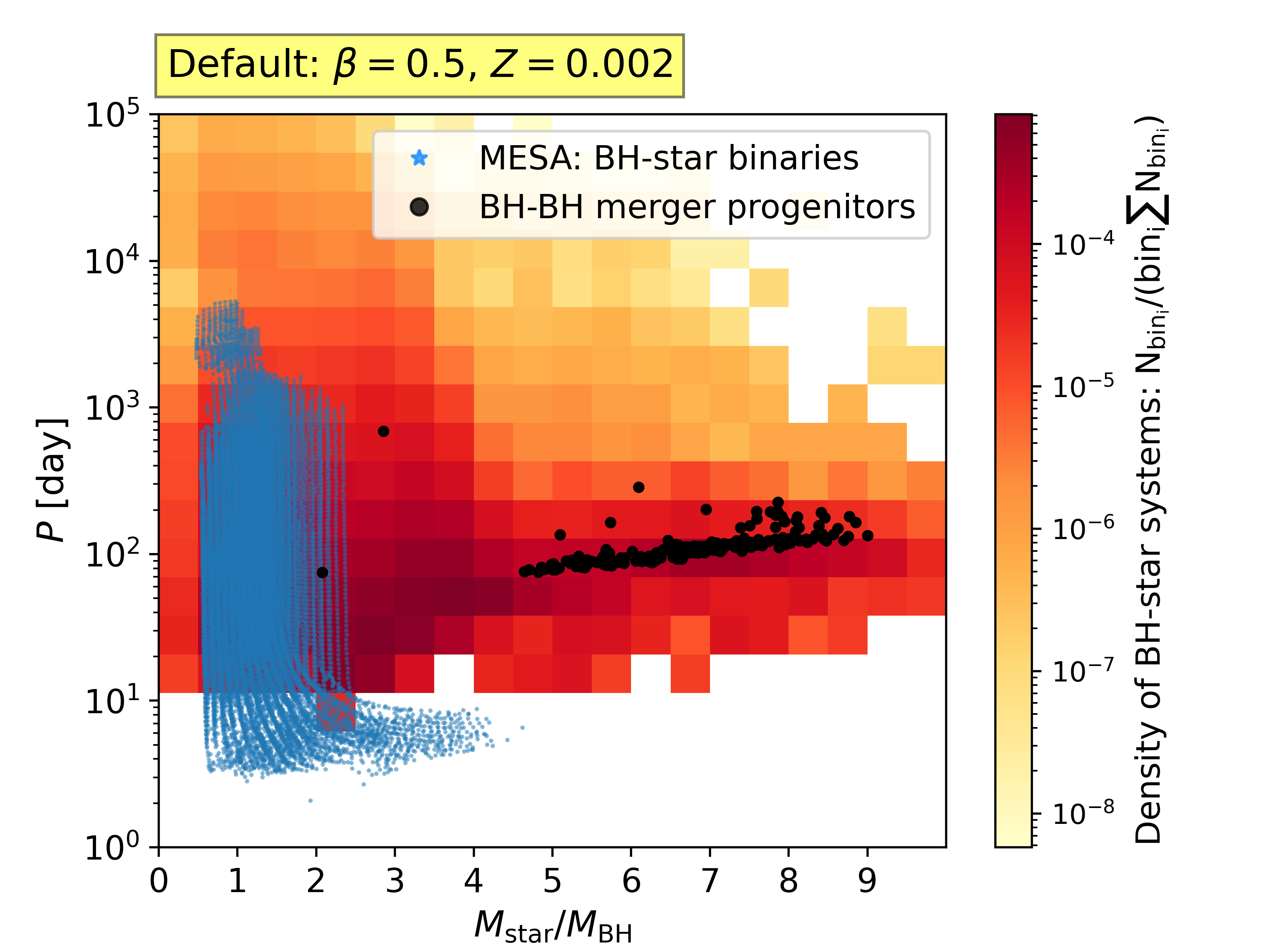}
\includegraphics[width=0.5\textwidth]{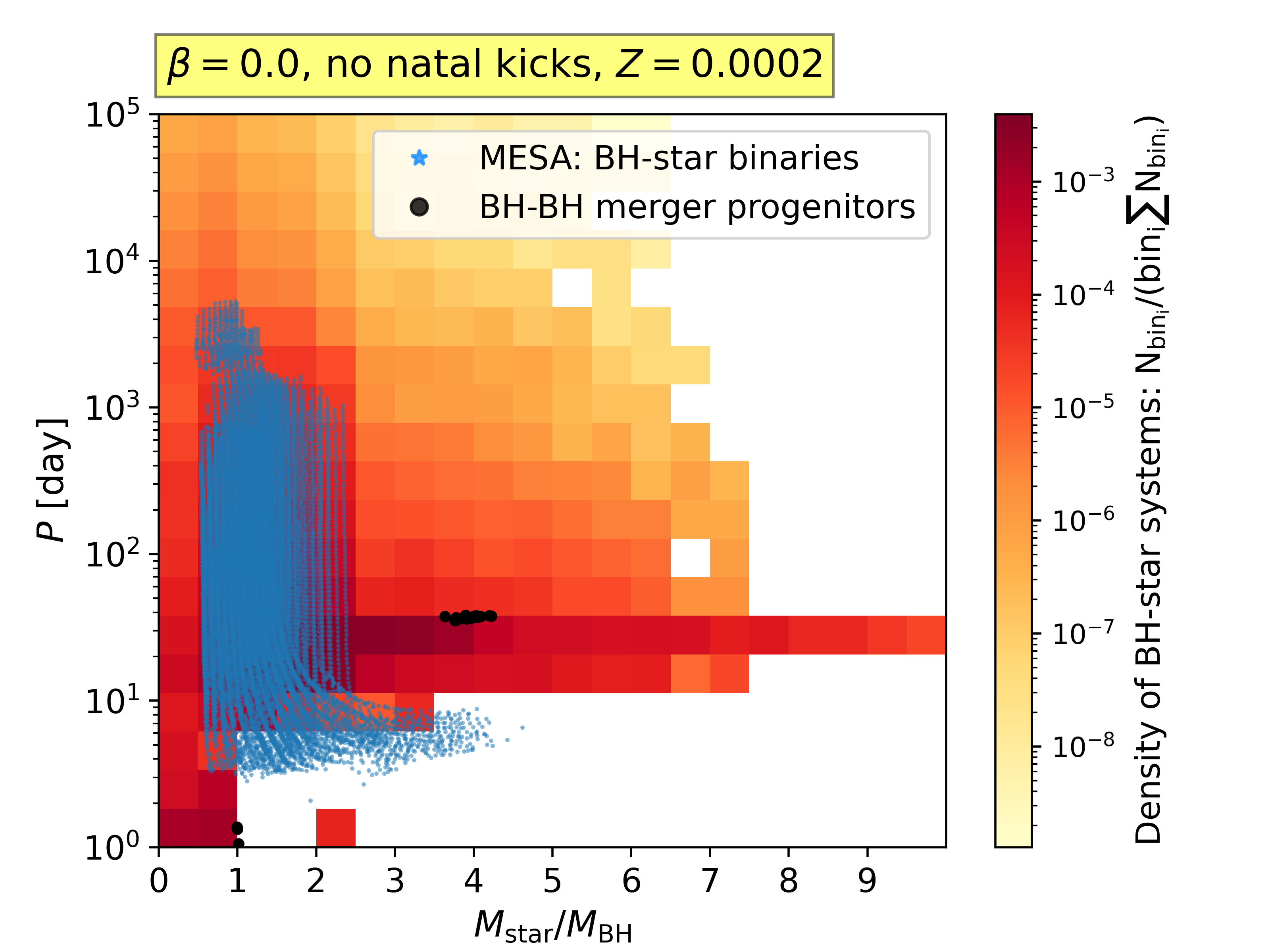}
\includegraphics[width=0.5\textwidth]{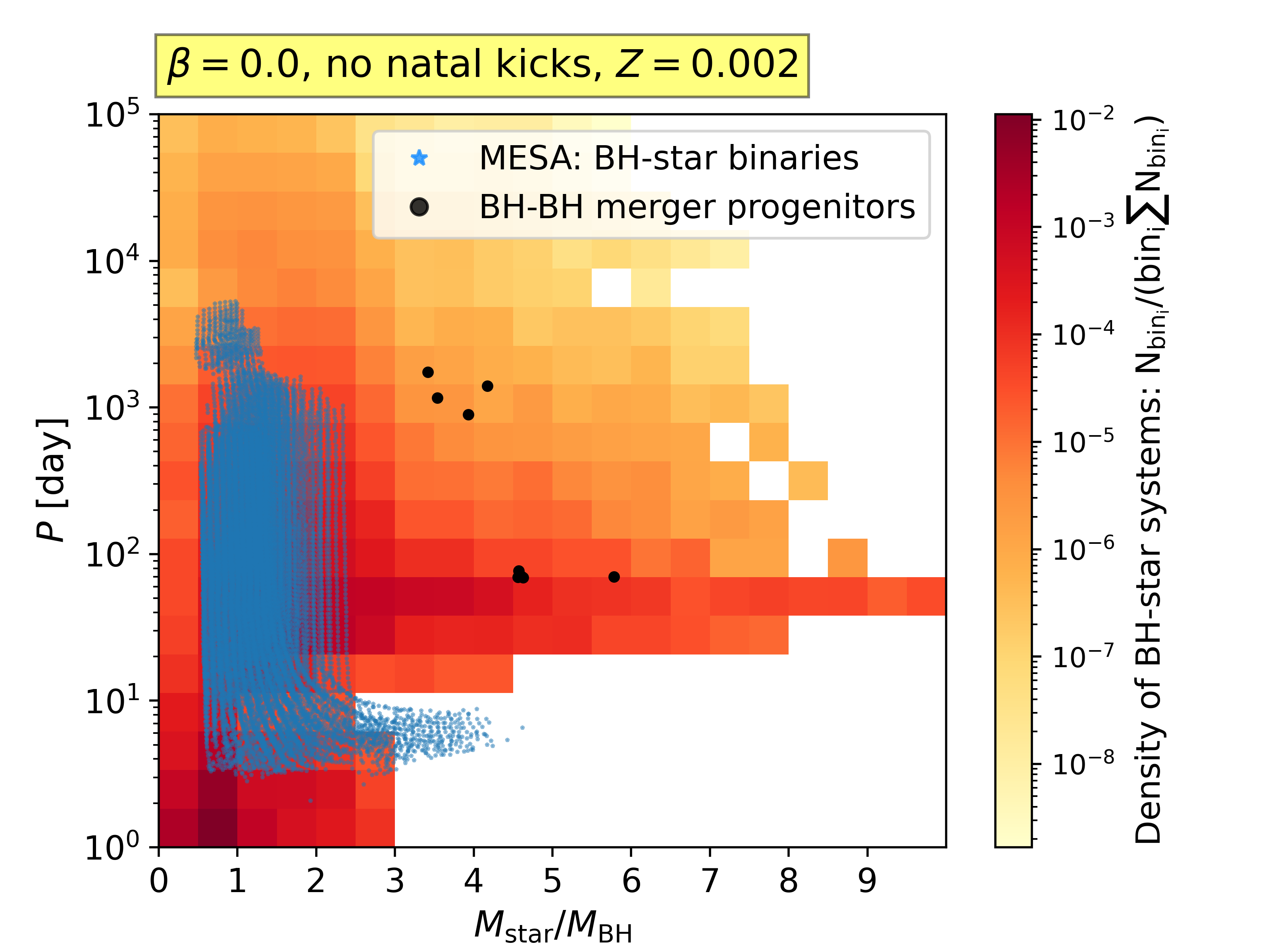}
\caption{Density grids of BH-star binaries using our rapid population synthesis code {\tt StarTrack} - red color; and detailed stellar evolution code {\tt MESA} from \cite{Xuinprep} - blue color. Black points are {\tt StarTrack} BH-star systems that would later evolve into BH-BH mergers via stable mass transfer subchannel. The upper panels show the grid for our default {\tt StarTrack} model. The bottom panels are results for the {\tt StarTrack} model with 100\% non-conservative mass transfer and no natal kick due to asymmetric mass ejection (except neutrino emission). The left panels are for {\tt StarTrack} models with 1\% solar metallicity, the right panels are for 10\% solar metallicity. {Detailed grid is for 10\% solar metallicity}. }
\label{fig: Detailed_models}
\end{figure*}

Figure \ref{fig: Detailed_models} demonstrates that once adopting in {\tt StarTrack} similar physical input assumptions to the ones in detailed stellar codes, the grids of BH-star binaries begin to overlap well. The panels of the Figure present a comparison between the detailed model grid for the 10\% solar metallicity (blue color) and different versions of {\tt StarTrack} grids (red color). The left and right panels correspond to the 1\% and 10\% solar metallicity {\tt StarTrack} grids, respectively. 

The upper panels of Fig \ref{fig: Detailed_models} show results for our default model which adopts relatively efficient mass accretion on the stellar companion ($\beta = 50\%$) during the first mass transfer phase and the standard model for natal BH kicks (fallback-decreased BH kicks). Such a set of assumptions allows for a common mass ratio reversal during the first RLOF which allows for the formation of unequal mass BH-star systems with moderate orbital periods. The mass transfer continues after the donor (initially more massive) is less massive than the companion (initially less massive) and during that phase, the binary orbit widens, suppressing the formation of very close BH-star binaries. The orbital period can be further widened by a BH natal kick.
 
The two bottom panels present results for a model with non-efficient mass transfer ($\beta = 0$) and no BH natal kicks ($\sigma = 0 \kms$) \footnote{The final BH-star orbits are still mildly affected by the neutrino emission, which in our models equal to 1\% of pre-core collapse mass of the BH progenitor.}. Such an input setup is calibrated to match assumptions adopted for the detailed grid. {\tt StarTrack} with the tuned assumptions produces equal-mass BH-star binaries with short orbital periods of $P \lessapprox 5$ days, in agreement with the results for detailed codes.

The effect of metallicity on the populations of BH-star systems produced by StarTrack is complex. In the case of relatively conservative mass transfer with $\beta=0.5$ and standard BH natal kicks (upper panels of Fig. \ref{fig: Detailed_models}) the orbital periods of the formed BH-star binaries are systematically wider in high metallicity ($Z=10 Z_{\odot}\%$) than in low metallicity ($Z = 1\% Z_{\odot}$). The separation of those systems is already quite wide due to mass ratio reversal during the first RLOF. In high metallicity, the orbits are additionally widened by strong stellar winds and high BH natal kicks (as in default model kicks tend to be inversely proportional to the BH mass). 

The model with non-efficient accretion during the first RLOF ($\beta = 0$) and no natal BH kicks (lower panels of Fig. \ref{fig: Detailed_models}) produces a fraction of BH-star systems with nearly equal BH-star masses with their low orbital periods $P \leq$5 days in both metallicities. The accretion limited to zero allowed some of the systems to avoid mass ratio reversal and remain on the close orbits after the first RLOF. The mass ratios of those close BH-star systems are less equal in high metallicity as the final BH mass is reduced by strong stellar winds. This grid is also in good agreement with the detailed grid, which is also for the systems with $Z \approx 10\%$.

\color{black}

\subsection{Other major uncertainties}

We would like to point out that the presented results are subject to several other uncertainties apart from the ones tested and described in the earlier sections. Here, we select the most relevant ones, that could have the most significant effect on the final distribution of mass ratios and spins of our population of BH-BH mergers.

Adopted efficiency of tidal spin-up of the helium cores by a BH companion, and therefore the fraction of highly spinning second-born BH, might be treated rather as an upper limit. A few factors, not considered in our models, may significantly decrease the number of highly spinning BH-BH mergers. First, after the mass transfer, we assume that the donor is completely stripped of its hydrogen envelope. The star may retain some fraction of the hydrogen layer, which later could affect its radius and the fate of the tight binary system. In particular, some of the stripped stars might experience further expansion at the later evolutionary stage and initiate the next phase of (stable or unstable) mass transfer, see \cite{2020Laplace} and \cite{Klencki2022}.  Second, our adopted, simplified approach does not consider possibly more sophisticated dependency on the metallicity and stellar winds that could weaken tidal interactions by taking away the star's angular momentum as well as by widening the orbit \citep{Detmers2008}. 

The increased amount of angular momentum loss during ongoing mass transfer could also affect its stability. In particular, it could destabilize mass transfer leading to higher mass transfer rates, and possibly a dynamically unstable mass transfer phase that ends with a successful CE or a merger \citep{Willcox2023}. This effect could also prevent the formation of some BH-BH mergers, especially in the model with increased angular momentum loss $\gamma = \gamma_{\rm L_{2}}$. Although many BH-star systems merge anyway due to efficient orbital contraction (and too small separation to fit the binary components), that effect might not be fully included in our simulations.

Another important assumption in our simulations is efficient circularization and synchronization due to tides (see Sec. \ref{sec:method}, so the systems always enter the RLOF phase on a circular orbit \citep{Hurley_2002,Belczynski2008}. The efficiency of tides is however highly uncertain and depends on the stellar structure \citep{Zahn1977,Zahn1989}. Note that some observations indicate that binaries could maintain some eccentricity even after the mass transfer phase \citep{Latham2007, Eldridge2009}. In our study, the assumption of efficient circularization may especially affect predictions for the model with high BH natal kicks. While in our default (fallback-decreased) model BHs are born with rather low natal kicks and mild eccentricities, the model with full BH natal kicks results in a significant fraction of highly eccentric systems ($e\gtrsim0.5$). The uncertain tidal circularisation affects especially the second RLOF phase. After the first BH was formed, the second mass transfer phase could be initiated once the orbit is still eccentric (see formation scenario in Fig. \ref{fig: neg_eff_spin}). Several studies tested different approaches to follow the evolution of orbital parameters and mass transfer rates during eccentric RLOF \citep[see e.g.][]{Davis2013,Dosopoulou2016a,Dosopoulou2016b,Hamers2019} and found it dependent on the system mass ratio. In the case of typical BH-BH merger progenitors in our stable mass transfer channel, donors are several times more massive than the accretors at the onset of the second RLOF. For such a mass-ratio regime eccentricity is expected to decrease or remain constant during the ongoing mass transfer phase \citep{Dosopoulou2016b, Hamers2019} and the orbital separation is expected to shrink, similarly to or more rapidly than in the case of circular RLOF \citep{Hamers2019}. Eccentricity at the onset of RLOF could also affect the stability of mass transfer. If the orbital separation shrinks more rapidly in the eccentric case than in the circular one, including eccentricity in our models would possibly lead to more common unstable mass transfer. On the other hand, for radiative-envelope donors, such as our BH-BH merger progenitors in the stable mass transfer scenario, it is expected that eventual instabilities would develop with time delay, possibly allowing binary orbit to circularize in the meantime due to tides \citep[see e.g.][]{Blagorodnova2021}. Mass transferring binaries with convective-envelope donors could instead initiate CE on eccentric orbits, which would affect the course of the phase and the final outcome \citep[see e.g.][]{Glanz2021}. 

We also point out that our mass transfer stability criteria are approximated by the results for a limited grid of massive binaries and metallicities studied in \cite{Pavlovskii2017}. The condition for whether mass transfer in the given system is stable or unstable could be much more complex and sensitive to several other factors like eccentricity or individual properties of the binary such as chemical composition and evolutionary stage of donor \citep[e.g.][]{Ge2023,Picco2024}. Independent origin of the (anti)correlation between effective spins and mass ratios ($\chi_{\rm eff}$ vs $q$) of BH-BH mergers formed in isolated binary evolution is the subject of Klencki et al. in prep.

\section{Conclusions} \label{sec:conclusions}

In this study, we show how isolated binary evolution could reproduce the possible anti-correlation between effective spins and mass ratios ($\chi_{\rm eff}$ vs $q$) reported for BH-BH mergers \citep[see e.g.][]{LIGOfullO3population2021, Callister2021}. We find that our model with revised mass transfer stability criteria limiting CE development and combined with low angular momentum loss during non-conservative mass transfer produces a broad peak of unequal-mass BH-BH mergers with mass ratios between $q \in (0.4-0.7)$. Once allowing for an efficient tidal spin up in the close BH-stripped helium core binaries, a significant fraction of those BH-BH systems merge with relatively high effective spins of $\chi_{\rm eff}>0.2$. The unequal-mass BH-BH mergers are produced by formation scenario including a mass ratio reversal during the first, relatively conservative mass transfer, and a low angular momentum loss during the second, highly non-conservative mass transfer phase. Such a mass ratio reversal scenario as well as a significant contribution of unequal mass ratio BH-BH mergers is consistent with some of the recent analyses of GW detections \citep{Adamcewicz2023, Rinaldi2023,Sadiq2023,Adamcewicz2024}. In contrast, the minor fraction of BH-BH mergers produced via a CE phase during the second RLOF tends to have rather equal masses and low spins. Therefore, the combination of the two formation subchannels: the dominant --stable mass transfer channel and the subdominant --CE channel may constitute a good match for the detected GW systems \citep{Neijssel2019,Bavera2021,Olejak2021a,Gallegos-Garcia2021,vanSon2022, Dorozsmai2024}.

However, BH-BH mergers' properties are sensitive to uncertain angular momentum loss during the non-conservative mass transfer phase. In particular, the tendency towards the formation of unequal-mass BH-BH mergers in stable mass transfer formation scenarios is no longer valid once we adopt more efficient angular momentum loss. Once we allow for high angular momentum loss, mass ratios of BH-BH mergers become much more diverse. The mass ratio distribution of BH-BH mergers is also affected by core-collapse physics, especially the assumption on BH natal kick magnitudes. High and mass-independent BH natal kicks lead to the disruption of many systems but also increase the relative fraction of equal-mass binaries among the population of BH-BH mergers due to high eccentricities. This study aims to present general trends and their sensitivity to uncertain physical assumptions. We did not try to calibrate our models to best match the detected BH-BH population. Note that the fraction of unequal-mass, highly-spinning BH-BH mergers in our default model could be overestimated due to e.g. the efficient tidal spin-up of helium core and a few assumptions favoring production of unequal BH-BH mergers via stable mass transfer (see discussion in Sec. \ref{sec:discussion}).

In this study, we also make a brief comparison of our grid of BH-star binaries with the results of detailed stellar evolution codes. We find that once adopting in rapid population synthesis code {\tt StarTrack} equivalent physical input assumptions regarding mass transfer efficiency and BH natal kicks, our grid of BH-star binaries overlap to a reasonable extent with the one generated using detailed stellar evolution code {\tt MESA}. Low mass transfer efficiency, however, reduces the fraction of highly unequal BH-star systems (${M}_{\rm star}/{M}_{\rm BH} \geq 4$), which are the main BH-BH mergers progenitors in our stable mass transfer scenario.  \\

Based on our results, we conclude that:
\begin{enumerate}
\item In contrast to the view widely spread in the GW community, isolated binary evolution does not necessarily favor the formation of BH-BH mergers with equal-mass components. A significant fraction of unequal-mass BH-BH mergers among the future GW detections could indicate the contribution of the isolated binary formation scenario via stable mass transfer. 
\item Population of BH-BH mergers with the broad peak in mass ratio distribution between $q \in 0.4-0.7$ would be consistent with the mass ratio reversal scenario during the first, relatively conservative mass transfer, and low angular momentum loss during the second, highly non-conservative mass transfer phase. The detection of BH-BH mergers dominated by equal mass-ratio systems is however consistent with both stable mass transfer channels (e.g. with higher angular momentum loss or high BH natal kicks) and classical CE channels ($\alpha$ formalism) which favors BH formation via direct core collapse. 
\item Future observations of massive binaries of a BH (or evolved star) and main sequence star with a highly unequal mass ratio would imply an efficient mass transfer phase during the first RLOF. However, observations of tight ($P \leq$5 days) and equal-mass BH-star systems would challenge the models with a relatively conservative mass transfer phase.
\end{enumerate}

\begin{acknowledgements}
AO and CW acknowledge funding from the Netherlands Organisation for Scientific Research (NWO), as part of the Vidi research program BinWaves (project number 639.042.728, PI: de Mink). AO acknowledges support from the Foundation for Polish Science (FNP) through the program
START 2022 and a scholarship from the Minister of Education and Science (Poland; 2022, 17th edition). AO would like to thank Selma de Mink, Lieke van Son, Daniel Holz, Ilya Mandel, Stephen Justham, Jakob Stegmann, Riccardo Buscicchio, and the reviewer for their useful comments. Sadly, while the work on the article was still in progress, Krzysztof Belczynski passed away. AO would like to thank him for the years of cooperation. JPL lost a friend and collaborator.
\end{acknowledgements}

\section{Data availability}
Data is available on request. The relevant fragments of the code and simulation results underlying this article will be shared on a request sent to the corresponding author: aolejak@mpa-garching.mpg.de.

\bibliographystyle{aa}
\bibliography{library}

\appendix
\subsection{The role of mass transfer instability in close orbit systems} \label{Sec: exp_instab}

 This subsection is devoted to the impact of mass transfer instability that might occur in close binaries or orbit systems. It may emerge once the donor star, which is still on its main sequence or early after, responds to the loss of the outer envelope with rapid expansion of deep layers, constantly increasing the relative difference between its radius and its Roche Lobe radius. The expansion of internal layers in this type of tight binary happens relatively close to the inner Lagrange point which accelerates the overflow and could lead to instability. The parameter space in terms of the donor masses and radius, for which binary systems experience this type of instability in our models is selected based on the results of \cite{Pavlovskii2017}. The detailed description of our revised mass transfer stability criteria together with the mass transfer stability diagram can be found in Section 3.1 and Figure 2 in \cite{Olejak2021a}.

\begin{figure*}[!htbp] 
\includegraphics[width=0.47\textwidth]{P_q_0.0002_SMT_inv_q.png}
\includegraphics[width=0.47\textwidth]{P_q_0.002_SMT_inv_q.png}\\
\includegraphics[width=0.47\textwidth]{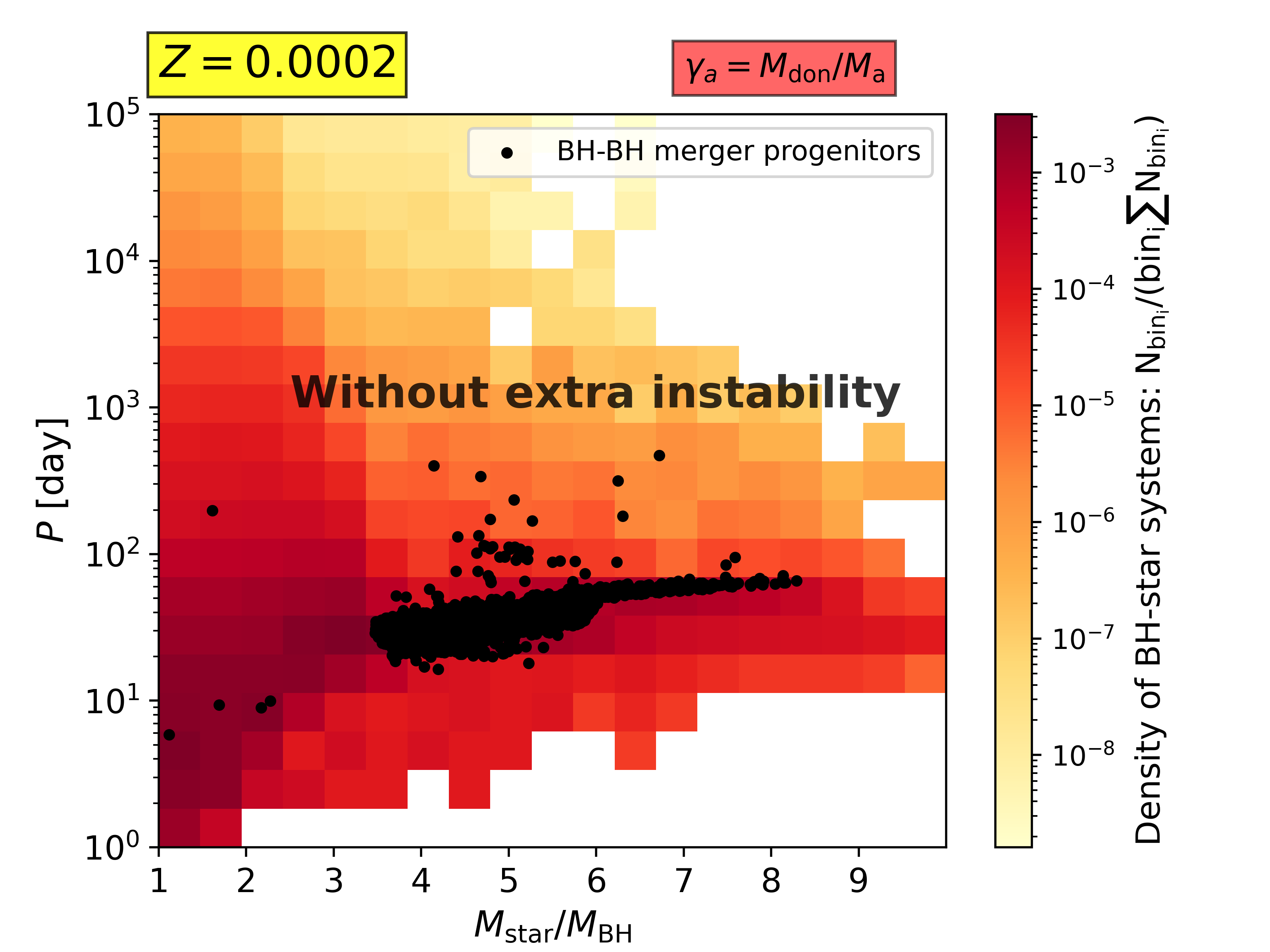}
\includegraphics[width=0.47\textwidth]{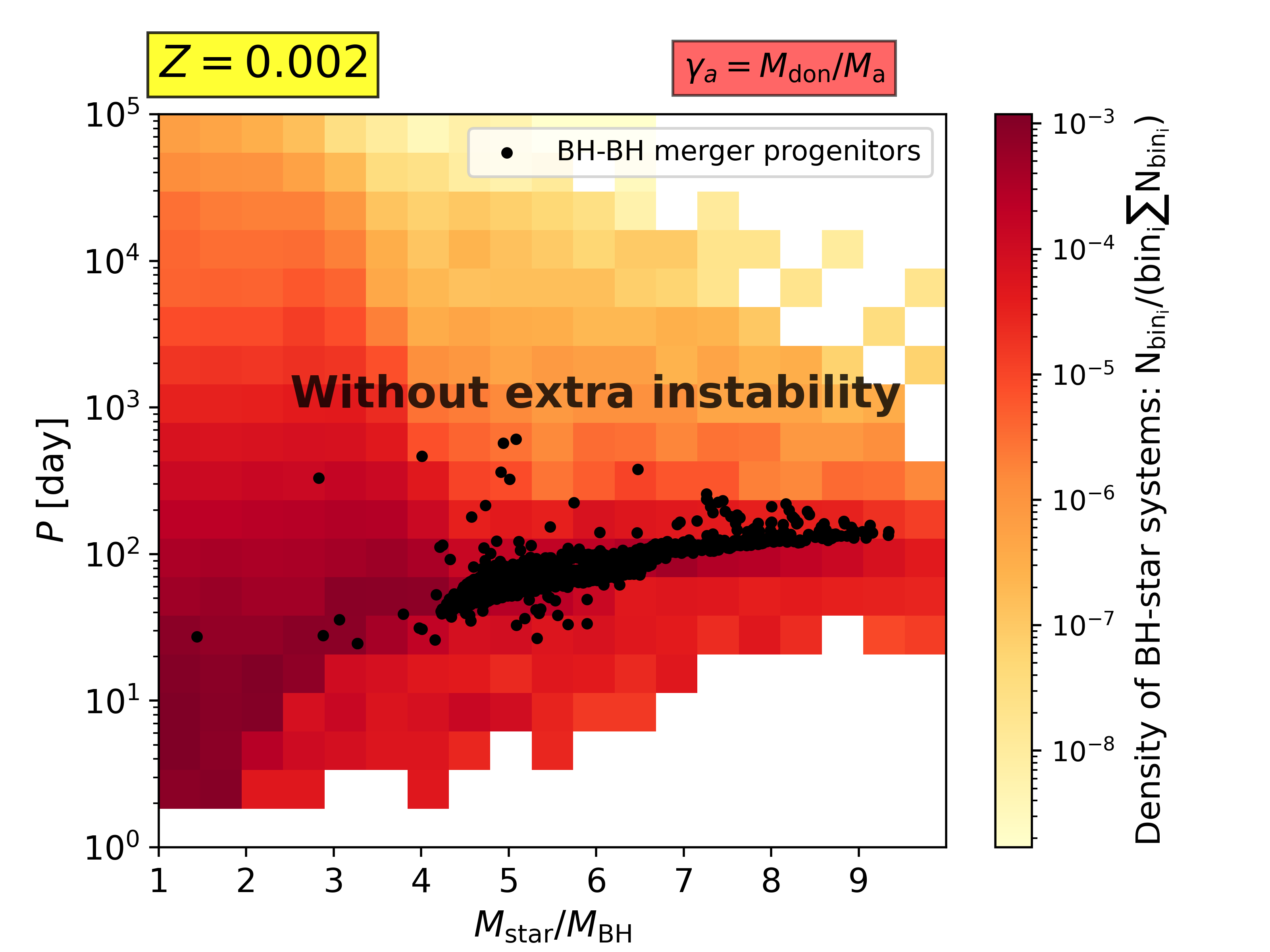}\\
\caption{Density plot of BH binaries with non-compact companion together with BH-BH progenitors (black points) in a grid of the system mass ratio and orbital period. Results only for the stable mass transfer subchannel. Left panels systems with $Z$ = 1\% $Z_{\odot}$, right panels with $Z$ = 10\% $Z_{\odot}$. The top two panels are with BH-star predicted for our default model, which includes the extra instability for tight BH-star binaries. The bottom panels are BH-star binaries formed in the model without the extra instability, which allows tight systems to proceed with stable mass transfer instead. This model, in contrast to our default one, produces a fraction of tight BH-star binaries ($P <5$days) and equal mass ratios. It also allows some of those close BH-star systems to later evolve into BH-BH mergers. Results only for stable mass transfer channels, without systems that evolved (or would evolve) through a common envelope.}
\label{fig: P_q_BH_MS_instability}
\end{figure*}

The emergence of this type of instability in mass-transferring binary star systems impacts the formation of tight BH-star systems. In particular, we find that once we include it in our simulations, we eliminate from our grids close BH-star systems with orbital periods $P<5$ days.  Those binary systems are expected to merge during the unstable mass transfer phase. That, in turn, also limits the number of possible BH-BH merger progenitors. However, the strong preference towards unequal-mass BH-star systems as BH-BH merger progenitors is not affected by including or excluding this type of instability in our models. Those results are demonstrated by the density grid of BH-star binaries and BH-BH progenitors in Figure \ref{fig: P_q_BH_MS_instability}. At the top panels, we present BH-star binaries formed in our default model, which includes an extra instability. Black points stand for binaries that later evolve into BH-BH mergers. The bottom panels show the results of the model in which the same systems go through for stable mass transfer evolution instead. This model, in contrast to the default one, allows for the formation of equal and moderate mass-ratio BH-star binaries $M_{\rm star}/M_{M_{\rm BH}}\leq 3$ on tight orbits $P<5-10$ days. The presented results are for two metallicities: 1\% solar metallicity (on the left) and 10\% solar metallicity (on the right). Note that efficient mass transfer on the companion and reversal of mass ratio during the first, stable mass transfer phase prevents the formation of close, highly unequal-mass BH-star binaries with $M_{\rm star}/M_{M_{\rm BH}}\geq 5$ no matter if we include an extra instability or not. The impact on BH-BH merger progenitors is visible on the plots. Model without instability slightly increases parameter space for BH-star systems that later evolve into GW sources. However, in both models, BH-BH mergers originate mainly from highly unequal-mass BH-star systems.

\end{document}